\documentclass[%
 reprint,
superscriptaddress,
 amsmath,amssymb,
 aps,
pra,
]{revtex4-2}

\pdfoutput=1
\usepackage{graphicx}
\usepackage{dcolumn}
\usepackage{bm}
\usepackage{hyperref}
\usepackage{braket}

\hypersetup{
    colorlinks=true,
    linkcolor=blue,
    citecolor = blue,
    filecolor=magenta,      
    urlcolor=blue,
    pdftitle={Fluxonium scaling},
    pdfpagemode=FullScreen,
    }

\usepackage{multirow}
\begin{document}

\preprint{APS/123-QED}

\title{Scalable High-Performance Fluxonium Quantum Processor}

\author{Long B. Nguyen}
\email{lbnguyen@lbl.gov}
\affiliation{Computational Research Division, Lawrence Berkeley National Laboratory, Berkeley, California 94720, USA}
\affiliation{Department of Physics, University of California, Berkeley, California 94720, USA}

\author{Gerwin Koolstra}
\affiliation{Computational Research Division, Lawrence Berkeley National Laboratory, Berkeley, California 94720, USA}
\affiliation{Department of Physics, University of California, Berkeley, California 94720, USA}

\author{Yosep Kim}
\altaffiliation{Current address: Center for Quantum Information, Korea Institute of Science and Technology (KIST), Seoul 02792, Korea}
\affiliation{Computational Research Division, Lawrence Berkeley National Laboratory, Berkeley, California 94720, USA}
\affiliation{Department of Physics, University of California, Berkeley, California 94720, USA}

\author{Alexis Morvan}
\altaffiliation{Current address: Google Quantum AI, Mountain View, CA 94043, USA}
\affiliation{Computational Research Division, Lawrence Berkeley National Laboratory, Berkeley, California 94720, USA}
\affiliation{Department of Physics, University of California, Berkeley, California 94720, USA}

\author{Trevor Chistolini}
\affiliation{Department of Physics, University of California, Berkeley, California 94720, USA}

\author{Shraddha Singh}
\affiliation{Department of Applied Physics, Yale University, New Haven, Connecticut 06511, USA}
\affiliation{Yale Quantum Institute, Yale University, New Haven, Connecticut 06511, USA}

\author{Konstantin N. Nesterov}
\affiliation{Bleximo Corp., Berkeley, California 94720, USA}

\author{Christian Jünger}
\affiliation{Computational Research Division, Lawrence Berkeley National Laboratory, Berkeley, California 94720, USA}
\affiliation{Department of Physics, University of California, Berkeley, California 94720, USA}

\author{Larry Chen}
\affiliation{Department of Physics, University of California, Berkeley, California 94720, USA}

\author{Zahra Pedramrazi}
\affiliation{Computational Research Division, Lawrence Berkeley National Laboratory, Berkeley, California 94720, USA}
\affiliation{Department of Physics, University of California, Berkeley, California 94720, USA}

\author{Bradley K. Mitchell}
\affiliation{Computational Research Division, Lawrence Berkeley National Laboratory, Berkeley, California 94720, USA}
\affiliation{Department of Physics, University of California, Berkeley, California 94720, USA}


\author{John Mark Kreikebaum}
\altaffiliation{Current address: Google Quantum AI, Mountain View, CA 94043, USA}
\affiliation{Department of Physics, University of California, Berkeley, California 94720, USA}
\affiliation{Materials Science Division, Lawrence Berkeley National Laboratory, Berkeley, California 94720, USA}


\author{Shruti Puri}
\affiliation{Department of Applied Physics, Yale University, New Haven, Connecticut 06511, USA}
\affiliation{Yale Quantum Institute, Yale University, New Haven, Connecticut 06511, USA}

\author{David I. Santiago}
\affiliation{Computational Research Division, Lawrence Berkeley National Laboratory, Berkeley, California 94720, USA}
\affiliation{Department of Physics, University of California, Berkeley, California 94720, USA}

\author{Irfan Siddiqi}
\affiliation{Computational Research Division, Lawrence Berkeley National Laboratory, Berkeley, California 94720, USA}
\affiliation{Department of Physics, University of California, Berkeley, California 94720, USA}
\affiliation{Materials Science Division, Lawrence Berkeley National Laboratory, Berkeley, California 94720, USA}


\begin{abstract}
    The technological development of hardware heading toward universal fault-tolerant quantum computation requires a large-scale processing unit with high performance. While fluxonium qubits are promising with high coherence and large anharmonicity, their scalability has not been systematically explored. In this work, we propose a superconducting quantum information processor based on compact high-coherence fluxoniums with suppressed crosstalk, reduced design complexity, improved operational efficiency, high-fidelity gates, and resistance to parameter fluctuations. In this architecture, the qubits are readout dispersively using individual resonators connected to a common bus and manipulated via combined on-chip RF and DC control lines, both of which can be designed to have low crosstalk. A multi-path coupling approach enables exchange interactions between the high-coherence computational states and at the same time suppresses the spurious static $ZZ$ rate, leading to fast and high-fidelity entangling gates. We numerically investigate the cross resonance controlled-NOT and the differential AC-Stark controlled-Z operations, revealing low gate error for qubit-qubit detuning bandwidth of up to $1~\mathrm{GHz}$. Our study on frequency crowding indicates high fabrication yield for quantum processors consisting of over thousands of qubits. In addition, we estimate low resource overhead to suppress logical error rate using the XZZX surface code. These results promise a scalable quantum architecture with high performance for the pursuit of universal quantum computation.

\end{abstract}

\maketitle


\section{\label{sec:introduction}Introduction}

Quantum computation and quantum information processing has opened a new frontier in science and technology, with recent advances validating our understanding of the quantum world and revealing the potential for novel applications \cite{harrow2017quantum,montanaro2016quantum,biamonte2017quantum,altman2021quantum}. Among various platforms, superconducting qubits has emerged as a promising candidate for the implementation of fault-tolerant universal quantum computing  \cite{devoret2013superconducting}, with the power of quantum information processing  \cite{arute2019quantum,zhu2021quantum} and novel quantum error correction schemes \cite{ofek2016extending,hu2019quantum,campagne2020quantum, grimm2020stabilization,gertler2021protecting,andersen2020repeated,marques2021logical,ai2021exponential,krinner2021realizing,zhao2021realizing} having been demonstrated. In addition, quantum simulation \cite{las2014digital,salathe2015digital,barends2016digitized,colless2018computation,mcardle2019error,blok2021quantum,arute2020observation} and optimization algorithms \cite{kandala2017hardware,google2020hartree} have been implemented using noisy intermediate scale quantum (NISQ) devices \cite{hashim2020randomized,jurcevic2021demonstration}. 



Approaching large-scale universal quantum computation requires further suppression of errors resulting from control imprecision and decoherence due to unwanted interaction with the noisy environment \cite{fowler2012surface, preskill2018quantum,stilck2021limitations}. The theory of quantum error correction (QEC) provides a promising path to reach this goal \cite{shor1996fault,kitaev2003fault,dennis2002topological,campbell2017roads}. The essence of the QEC strategy is to encode quantum information in non-local entangled states such that local errors cannot corrupt it \cite{terhal2015quantum}. According to the threshold theorem, when the system is operated with errors below the \textit{accuracy threshold}, an arbitrarily good protection against decoherence can be achieved \cite{aharonov2008fault,knill1998resilient,preskill1998reliable}. In practice, the requirement to encode \textit{logical qubits} using a redundant number of \textit{physical qubits} imposes a large resource overhead that is challenging to achieve. Besides decoherence of physical qubits, crosstalk \cite{sarovar2020detecting,winick2021simulating}, frequency crowding \cite{brink2018device,hertzberg2021laser}, and leakage out of the computational subspace \cite{varbanov2020leakage,mcewen2021removing} are also the central problems to overcome upon scaling up. For example, the most popular QEC code for planar architecture, the surface code \cite{bravyi1998quantum,fowler2009high,wang2011surface,fowler2012surface}, has an accurate threshold of approximately $1\%$ and only requires nearest neighbor interactions, yet only small-distance codes have been recently explored in superconducting-circuit architectures based on transmon devices \cite{andersen2020repeated,ai2021exponential,marques2021logical,krinner2021realizing,zhao2021realizing}. 

One promising superconducting qubit in the quest toward constructing a fault-tolerant quantum computer is fluxonium \cite{manucharyan2009fluxonium}, due to its long coherence times and high anharmonicity. The circuit consists of three elements in parallel: a capacitor, a Josephson junction, and a superinductor. The inductive shunt eliminates the qubit's offset charge \cite{koch2009charging} and the large inductance value suppresses its sensitivity to flux noise. Fluxonium can be tuned \textit{in-situ} by threading an external magnetic flux through the circuit loop. When this flux bias is at a value equal to half-integer flux quantum, the $|0\rangle\rightarrow|1\rangle$ transition has low frequency, with $\omega_{01}/2\pi$ below 1 GHz, which helps slow down dielectric loss effect, leading to a long relaxation time $T_1$. At the same time, the circuit also becomes first-order insensitive to flux noise, resulting in long coherence times $T_2$ primarily limited by $T_1$. This operating regime is thereby referred to as the \textit{high-coherence regime} \cite{nguyen2019high}, with $T_2$ consistently in the range of a few hundred microseconds in both 3D \cite{nguyen2019high} and planar \cite{zhang2020fast} experiments. Recent advances in fabrication and cryogenic shielding have elevated the  relaxation time $T_1$ and subsequently the coherence time $T_2$ to over a millisecond in a 3D device \cite{somoroff2021millisecond}.

 In addition, fluxonium's high anharmonicity can be exploited to operate high-fidelity single-qubit operations, with microwave control error of $\sim 10^{-4}$ using an $80~\mathrm{ns}$-long pulse \cite{somoroff2021millisecond}, and fast flux control error of $\sim 10^{-3}$ using a $20~\mathrm{ns}$-long pulse \cite{zhang2020fast}. High readout fidelity with error $\sim 1\%$ using over a hundred of resonator photons has been demonstrated \cite{gusenkova2021quantum,takmakov2021minimizing}, while qubit reset with error $\sim 3\%$ using microwave \cite{zhang2020fast} or $\sim 1\%$ using active feedback \cite{gebauer2020state,gusenkova2021quantum} have been realized. The first experimental implementations of two-qubit gates were based on interaction between $|1\rangle\rightarrow |2\rangle$ transitions of two capacitively coupled fluxoniums housed in a 3D copper cavity, where a controlled-phase gate is induced by a microwave pulse applied near these transitions \cite{nesterov2018microwave}. The experimental implementation showed gate fidelity as high as $0.992$, limited by the coherence times of the participating non-computational states, which are unprotected \cite{ficheux2021fast,xiong2021arbitrary}. Recently, an iSWAP gate \cite{chen2021fast} between two planar capacitively-coupled fluxoniums was realized by tuning their computational transitions $\omega_{01}$'s to be on-resonant using fast flux, with corresponding gate fidelity as high as $0.997$ \cite{bao2021fluxonium}. The performance of this gate scheme is intrinsically limited by the lower coherence time away from the flux sweet spot, and its operation may involve additional complications such as spectator errors in large-scale devices. Currently, there is no clear approach to scale up fluxonium devices with projected performance to surpass state-of-the-art architectures.

\begin{figure}[ht!]
        \includegraphics[width=0.5\textwidth]{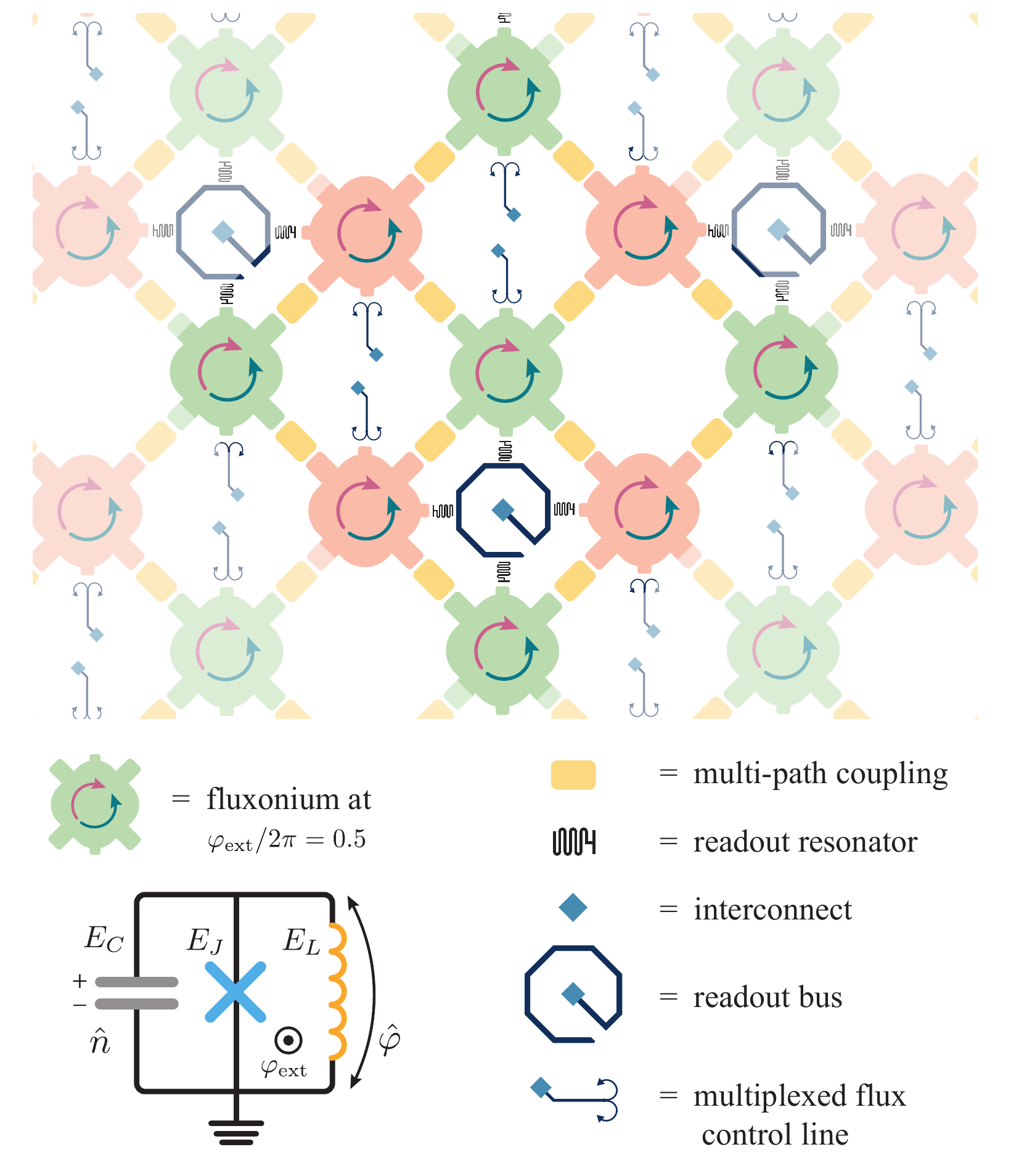}
        \caption{\label{fig0}Fluxonium quantum processor schematic. The qubits are fluxonium circuits biased at half-integer flux quantum, and their lowest eigenstates are symmetric and anti-symmetric superpositions of fluxon states corresponding to supercurrents flowing clockwise and anti-clockwise in the circuits. The two different colors represent data and ancilla qubits in a surface code lattice. The qubits are readout dispersively via individual coplanar waveguide resonators, four of which are capacitively connected to a common superconducting bus. The diplexed flux lines provide both DC bias and RF controls. These components are connected to electronics instruments via microwave interconnects. The qubit-qubit interaction follows a multi-path coupling approach, which allows fast and high-fidelity entangling gates.}
\end{figure}

In this work, we propose a scalable high-performance quantum architecture based on fluxonium which can be used to implement resource-intensive algorithms and quantum error correction codes such as the surface code. First, we discuss the important properties of the qubit on which we develop and explain the proposed parameters as well as our approaches in the rest of the paper. We review the dispersive readout scheme which involves a coupled fluxonium-resonator system, pointing out the advantages in readout multiplexing with fluxoniums, such as the broad range for optimal readout resonator frequency. In addition, since the qubit frequency is far away from the readout frequency, Purcell filters \cite{reed2010fast,jeffrey2014fast,sete2015quantum,bronn2015broadband,walter2017rapid} are not required for fast measurements. Then, we show via numerical simulation that single-qubit microwave control can be done in as fast as ten nanoseconds with coherent error below $10^{-6}$, with flux driving performing better than charge driving. Together with the lack of Purcell filters, diplexing the RF flux control with DC bias will substantially reduce the design complexity of the chip.

The keystone in developing large-scale quantum devices is designing multi-qubit interactions. Since the computational states in fluxonium have long coherence times and high anharmonicity at the fixed half-integer flux quantum bias, the optimal approach is to engineer entangling gates based on the transverse coupling between these states. Meanwhile, it is also important to suppress the spurious static longitudinal $ZZ$ rate, which leads to an always-on entangling operation, and may expand into chip-scale correlated errors \cite{sundaresan2020reducing,sarovar2020detecting,winick2021simulating}. Here, we propose a multi-path coupling scheme which enables a fast direct exchange interaction rate between the computational states, with $J_\mathrm{eff}\sim 10~\mathrm{MHz}$, and at the same time negates the static $ZZ$ rate across a wide range of qubit parameters. We numerically demonstrate high-fidelity microwave-activated two-qubit gates based on the cross-resonance \cite{de2010selective,rigetti2010fully,chow2011simple,chow2012universal} and differential AC-Stark shift \cite{xiong2021arbitrary,mitchell2021hardware,wei2021quantum} effects, with coherent gate error as low as $10^{-6}$, and consistently below $10^{-2}$ across a wide range of qubit-qubit detuning. Since these gates only involve high-coherence computational states, gate error due to decoherence is small. Moreover, they can be implemented using the same control lines and microwave electronics for single-qubit gates, improving resource efficiency in scaling up.

Upon advancing to large-scale devices, frequency crowding is a major challenge to overcome. We investigate this problem by considering the constraints imposed by the gate simulations. Based on previous studies on Josephson junction nano-fabrication, we consider practical frequency dispersion in fluxonium qubits biased at the optimal external flux, and show that the probability of constructing a frequency-collision-free device is close to unity for large-scale chips consisting of over thousands of qubits arranged in a square lattice. In addition, we discuss further error suppression of this platform using the XZZX surface code \cite{ataides2021xzzx}, showing exponential reduction of logical error rate $\varepsilon_L$, with $\varepsilon_L\sim 10^{-7}$ for code distance $d=11$. Our results also indicate the importance of improving readout and initialization fidelity.

A schematic of the proposed architecture is shown in Fig.~\ref{fig0}. The paper is structured to explore each component of this platform as follows. We revisit important properties of the high-coherence fluxonium qubit in Sec.~\ref{sec:fluxonium}. In Sec.~\ref{sec:readout}, we discuss dispersive measurement of the qubit via an ancilla resonator, reviewing recent results and emphasizing important considerations upon scaling up. In Sec.~\ref{sec:single-qubit}, we explore single-qubit operations, comparing flux and charge microwave controls. We investigate different coupling schemes in Sec.~\ref{sec:multi-qubit}, revealing the advantage of the multi-path coupling approach. We numerically simulate the cross resonance controlled-NOT (CNOT) \cite{de2010selective,rigetti2010fully,chow2011simple}  and the differential AC-Stark controlled-Z (CZ) \cite{mitchell2021hardware,wei2021quantum} gates in Sec.~\ref{sec:two_qubit_gates}. After discussing experimentally feasible frequency dispersion of the computational states the frequency allocation constraints, we compute the expected fabrication yield of large-scale devices in Sec.~\ref{sec:scaling}. Motivated by these results, we simulate logical error suppression using the XZZX surface code \cite{ataides2021xzzx} in Sec.~\ref{sec:qec}. Finally, we summarize the main ideas of the paper in Sec.~\ref{sec:summary}.



\section{\label{sec:fluxonium}Fluxonium qubit}

\begin{figure*}[ht]
        \includegraphics[width=0.8\textwidth]{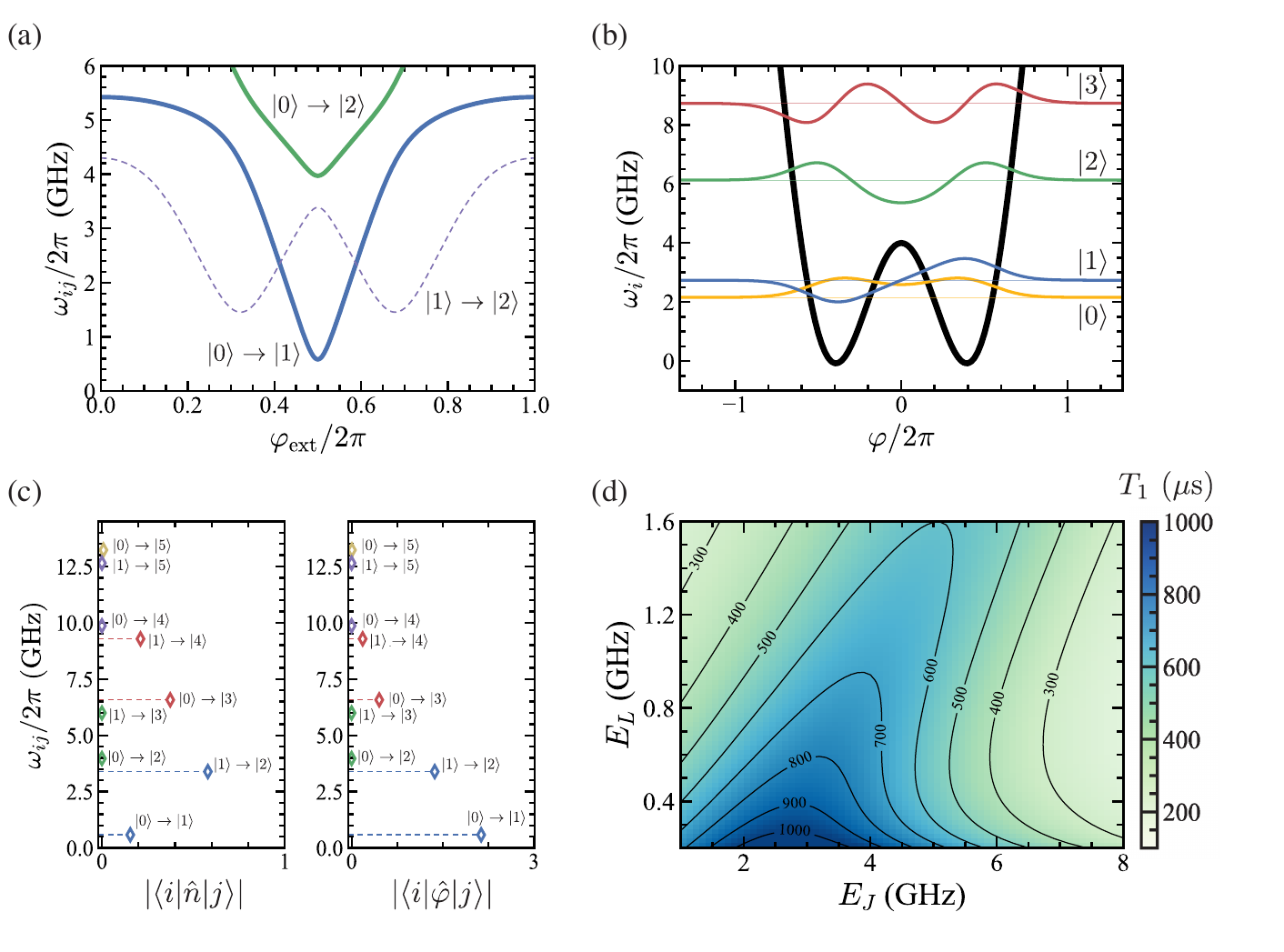}
        \caption{\label{fig1} Fluxonium properties. \textbf{(a)} Transition energy spectrum $\omega_{ij} \equiv \omega_{|j\rangle}-\omega_{|i\rangle}$ of a fluxonium circuit with $\{E_J, E_C, E_L \} = \{4,1,1 \} ~\mathrm{GHz}$. The solid lines are transitions from the ground state $|0\rangle$, and the dashed line is transition $|1\rangle\rightarrow |2\rangle$. \textbf{(b)} Potential well (black) and wave functions in the phase basis corresponding to the first four eigenstates at half-integer flux quantum bias $\varphi_\mathrm{ext}/2\pi=0.5$. 
        \textbf{(c)} Charge and phase matrix elements corresponding to transitions from $|0\rangle$ and $|1\rangle$ at $\varphi_\mathrm{ext}/2\pi=0.5$. The vertical axis shows the corresponding frequencies of these transitions. \textbf{(d)} Estimated energy relaxation times $T_1$ limited by dielectric loss with loss tangent $\tan \delta_\mathrm{diel} = 2\times 10^{-7}$ at 5 GHz, and quasiparticles in the inductor with normalized density $x_\mathrm{qp}=2\times 10^{-9}$. The charging energy is $E_C=1~\mathrm{GHz}$.}
    \end{figure*}
    
    A fluxonium circuit \cite{manucharyan2009fluxonium} consists of three elements in parallel (Fig.~\ref{fig0}, bottom left): a capacitor with charging energy $E_C=e^2/2C$, a Josephson junction with Josephson energy $E_J=I_c\phi_0$, and an inductor with inductive energy $E_L=\phi_0^2/L$, where we denote $I_c$ as the junction's critical current, and $\phi_0=\Phi_0/2\pi$ as the reduced flux quantum. The Hamiltonian of the system can be written as
        \begin{equation}
        \label{eqn:fluxonium_hamiltonian}
            \hat{H}/h = 4E_C\hat{n}^2 -E_J\cos(\hat{\varphi} + \varphi_\mathrm{ext}) + \frac{1}{2} E_L\hat{\varphi}^2 ,
        \end{equation}
    where $\hat{n}$ is associated with the number of excess Cooper pairs on the capacitive electrodes, and $\hat{\varphi}$ is the superconducting phase twist across the inductor. These two quantum mechanical operators obey the commutation relation $[\hat{\varphi},\hat{n}]=i$. In the harmonic oscillator basis with creation (annihilation) operator $\hat{b}^\dagger$ ($\hat{b}$), they can be written as 
        \begin{equation}
        \label{eqn:operators}
            \begin{split}
        \hat{\varphi} = \frac{1}{\sqrt{2}}\left(\frac{8E_C}{E_L}\right) ^ {\frac{1}{4}}(\hat{b}^\dagger + \hat{b}),\\
        \hat{n} = \frac{i}{\sqrt{2}}\left(\frac{E_L}{8E_C}\right) ^ {\frac{1}{4}}(\hat{b}^\dagger - \hat{b}),
            \end{split}
        \end{equation}
    such that the phase and charge operators resemble the position and momentum operators, respectively. In addition, their relative amplitudes generally depend on the energy ratio $E_C/E_L$.
    
    
    We can interpret the Hamiltonian in Eq.~(\ref{eqn:fluxonium_hamiltonian}) as describing the motion of a fictitious particle with kinetic energy proportional to $E_C$ in a potential landscape shaped by $E_J$ and $E_L$, where the ratio $E_J/E_L$ determines the number of local wells within the quadratic envelope. The potential can be further tuned by applying an external magnetic flux, written in normalized form as $\varphi_\mathrm{ext} = 2\pi \Phi_\mathrm{ext}/\Phi_0$. For the qubit to be in the fluxonium regime, the energy scales are usually designed to satisfy $2\leq \{E_J/E_L,E_J/E_C\}\leq 10$. The $E_J/E_L$ condition determines the number of local wells, and the $E_J/E_C$ condition corresponds to the tunneling rate across the wells.
    
    Typically, the inductive energy in fluxonium is $E_L\sim 0.5~\mathrm{GHz}$, corresponding to a large inductance of $L\sim 325~\mathrm{nH}$. To fulfill this stringent requirement, different strategies have been employed to implement superconducting nano-inductors with impedance surpassing the resistance quantum $R_Q\sim 6.5~\mathrm{k\Omega}$, commonly referred to as \textit{superinductors}. They are typically constructed based on the phenomenon of kinetic inductance, with experimental realizations including arrays of Josephson junctions \cite{manucharyan2009fluxonium,masluk2012microwave}, superconducting nanowires \cite{shearrow2018atomic,hazard2019nanowire,niepce2019high}, or disordered superconductors such as granular aluminum \cite{maleeva2018circuit,grunhaupt2019granular,kamenov2020granular}. Recently, geometric superinductors have also been successfully fabricated and characterized \cite{peruzzo2020surpassing,peruzzo2021geometric}.
    
    Besides shaping the potential landscape, the superinductor also helps protect the qubit from decoherence. On one hand, the inductive shunt makes the excess charge on the capacitive electrodes continuous \cite{koch2009charging}, nulling the qubit's sensitivity to charge offset. Since fluxonium is inherently insensitive to charge noise, the charging energy $E_C$ may take arbitrary values, and thus can be used as a flexible tuning knob. On the other hand, the large inductance suppresses the qubit's sensitivity to $1/f$ flux noise \cite{lin2018demonstration,nguyen2019high}. Within the scope of this work, we consider the superinductor as an ideal circuit element, assuming that phase slips \cite{pop2010measurement,manucharyan2012evidence,masluk2012microwave}, cross-Kerr \cite{maleeva2018circuit}, and collective modes effects \cite{viola2015collective} are negligible.
    
     Circuits with large $E_J/E_C$ ratio are called \textit{heavy fluxonium} \cite{lin2018demonstration, earnest2018realization}, with the unique property of having long-lived circulating current states called fluxons at certain external fluxes where their wave functions are localized. In this regime, the potential and subsequently the fluxon transition energies are linearly dependent on external flux. Hence, their coherence times $T_2$ are limited by $1/f$ flux noise to about $5~\mathrm{\mu s}$ \cite{lin2018demonstration}. Heavy fluxonium biased at the symmetric flux bias becomes insensitive to flux noise up to first order, and has been shown to have long coherence times \cite{nguyen2019high,zhang2020fast,somoroff2021millisecond}.
    
    In this work, we focus on a moderate parameters configuration with $\{E_J,E_C,E_L\}\sim \{4,1,1\}~\mathrm{GHz}$, which correspond to a compact qubit that can be reliably fabricated. In addition, these parameters offer key scalability advantages which we discuss throughout the paper. The transition energy spectrum of the circuit is shown in Fig.~\ref{fig1}(a), with transitions from the ground state $|0\rangle$ and first excited state $|1\rangle$ plotted in solid and dashed lines, respectively. Their relative differences indicate the anharmonicity of the system. Energy transitions near  $\varphi_\mathrm{ext} = 0$ are similar to those of a standard transmon qubit, allowing straightforward device characterization.
    
    When the circuit is biased at the half-integer flux sweet spot $\varphi_\mathrm{ext}/2\pi=0.5$, the two lowest eigenstates correspond to symmetric and anti-symmetric superpositions of persistent current states corresponding to supercurrents flowing in the clockwise and counter-clockwise directions in the circuit, similar to those found in flux qubits \cite{mooij1999josephson,orlando1999superconducting,friedman2000quantum,van2000quantum,chiorescu2003coherent}. The transition energy is then simply the splitting gap resulting from tunneling of a fictitious particle across the potential barrier, the amplitude of which depends strongly on $\sqrt{E_J/E_C}$ and linearly on $E_L$ \cite{manucharyan2009fluxonium}. For our proposed circuit parameters, this frequency is $\omega_{01}/2\pi= 0.58~\mathrm{GHz}$. Fig.~\ref{fig1}(b) shows the energy spectrum and wave functions of the first four states at the symmetric flux bias $\varphi_\mathrm{ext}/2\pi=0.5$. Since the quadratic potential well only splits into local ones at sufficiently low frequency, the higher states resemble Fock states in a harmonic oscillator, albeit with high nonlinearity. They are separated from the two lowest states by a large energy gap, with $\omega_{12}/2\pi= 3.39~\mathrm{GHz}$. The high anharmonicity helps alleviate state leakage and frequency crowding significantly, as we shall explore in later sections. Below, we refer to the computational subspace as the Hilbert space consisting of the $|0\rangle$ and $|1\rangle$ states.
    
    Selection rules in fluxonium can be exploited to engineer novel coupling schemes for quantum information processing purposes. The matrix elements at half-integer flux quantum bias are shown in Fig.~\ref{fig1}(c) and can be summarized as follows. For the computational states, since their wave functions have substantial overlap in the phase basis due to hybridization, the corresponding matrix element $\varphi_{01}\equiv\langle 0|\hat{\varphi} |1\rangle$ is high, while the charge matrix element $n_{01}\equiv\langle 0|\hat{n} |1\rangle$ is substantially smaller. For the non-computational states, the phase (charge) matrix elements are smaller (larger) than that of the computational states. As a result, charge coupling is preferable if interaction involving higher states is desired, and flux coupling is preferable if interaction involving the computational states is intended. For example, charge coupling was used in the microwave-activated two-qubit controlled-phase gate scheme involving $|1\rangle\rightarrow |2\rangle$ transition \cite{nesterov2018microwave,ficheux2021fast,xiong2021arbitrary}, and flux coupling resulted in strong hybridization of the lowest eigenstates in the fluxonium molecule \cite{kou2017fluxonium}. In general, charge matrix elements are smaller than phase matrix elements due to the large inductance in the circuit ($8E_C>E_L$ in Eq.~(\ref{eqn:operators})). 
    
    
   While the multi-level aspect of fluxonium can be utilized to engineer novel coupling schemes, certain transitions are forbidden, alleviating spectral crowding problems and suppressing coupling to high-frequency spurious modes. Notably, only transitions with an odd number of excitation are allowed at half-integer flux bias due to parity, $n_{ij} = \varphi_{ij}=0$ for all $(i+j)_\mathrm{even}$. This allows selective coupling to any odd transition without interacting with the nearby even transition. In addition, the high-frequency region of the spectrum resembles that of a harmonic oscillator, so higher-order transitions have vanishing matrix elements. Thus, we can neglect coupling between fluxonium and the surrounding environment beyond 10 GHz.
    
    The proposed parameters also correspond to long coherence times of the computational states at the half-integer flux bias. Since the coherence time $T_2$ is primarily limited by the relaxation time $T_1$ here \cite{nguyen2019high,zhang2020fast,somoroff2021millisecond}, we focus on analyzing $T_1$. To this end, we estimate the energy relaxation time $T_1$ for fixed charging energy $E_C = 1~\mathrm{GHz}$ and different $E_J$, $E_L$ values, using limits imposed by dielectric loss with $\tan \delta_\mathrm{diel} = 2\times 10^{-7}$, corresponding to a transmon with frequency $\omega_{01}/2\pi=5~\mathrm{GHz}$ having a relaxation time of $160~\mathrm{\mu s}$, and quasiparticles in the inductor with normalized density $x_\mathrm{qp}= 2\times 10^{-9}$, corresponding to a fluxonium qubit with parameters as reported in Ref.~\cite{somoroff2021millisecond} having relaxation time $T_1=1~\mathrm{ms}$ at absolute temperature (see Appendix~\ref{appendix:fluxonium}). The result in Fig.~\ref{fig1}(d) shows that while the quality of the surrounding environment may change the absolute $T_1$ limit, the parameter space around $E_J = 4~\mathrm{GHz}$, $E_L = 1~\mathrm{GHz}$ gives consistently high $T_1$. 
    
    In multi-qubit devices, we propose tuning the qubit frequency by varying the inductive energy $E_L$ from $0.5$ to $1.6$ GHz, corresponding to qubit frequency $\omega_{01}/2\pi$ in the range $[237-1163]~\mathrm{MHz}$. As shown in Fig.~\ref{fig1}(d), a circuit with $E_L$ varying in this range is expected to have high $T_1$. We note that although the region $E_J \sim [2-4]~\mathrm{GHz}$, $E_L<0.4~\mathrm{GHz}$ has the highest $T_1$, the corresponding qubit frequencies are quite low, and hence do not work well with our proposed gate operations. Besides, reliable fabrication of a small junction corresponding to $E_J\sim 2~\mathrm{GHz}$ will be a difficult technical challenge. Thus, we focus on varying qubit frequency by changing $E_L$ and keeping $E_J= 4~\mathrm{GHz}$ for scalability. In practice, all three parameters can be fine-tuned to vary the qubit frequency $\omega_{01}$. We emphasize the flexibility of our approach: fluxonium circuits with parameters slightly deviated from the ones proposed in this work can still be scaled up favorably following the principles discussed in the following sections. On the other hand, we believe it is valuable to explore different parameter regimes in the future.

\section{\label{sec:readout}Readout}
    Dispersive readout in circuit quantum electrodynamics (cQED) architectures offers key advantages in the quest to build quantum computers \cite{blais2007quantum,blais2021circuit}. Since the qubit's transition frequency is far away from that of the resonator, the probability of qubit excitation being converted to cavity photons is negligible. This detuning thus protects the qubit from Purcell decay \cite{purcell1946resonance,houck2008controlling}. Dispersive readout has also been demonstrated to be fast and to have high fidelity \cite{walter2017rapid}, even when multiple qubits are readout simultaneously in multiplexed fashion \cite{chen2012multiplexed,heinsoo2018rapid,kundu2019multiplexed}. Quantum non-demolition measurement also allows qubit initialization by heralding \cite{johnson2012heralded,riste2012initialization,salathe2018low}. It has thus become the standard technique in modern superconducting quantum information processors \cite{arute2019quantum,jurcevic2021demonstration,hashim2020randomized}. In this section, we review the dispersive interaction framework in a fluxonium-resonator system and discuss its advantages upon scaling up.
    
\begin{figure}
        \includegraphics[width=0.47\textwidth]{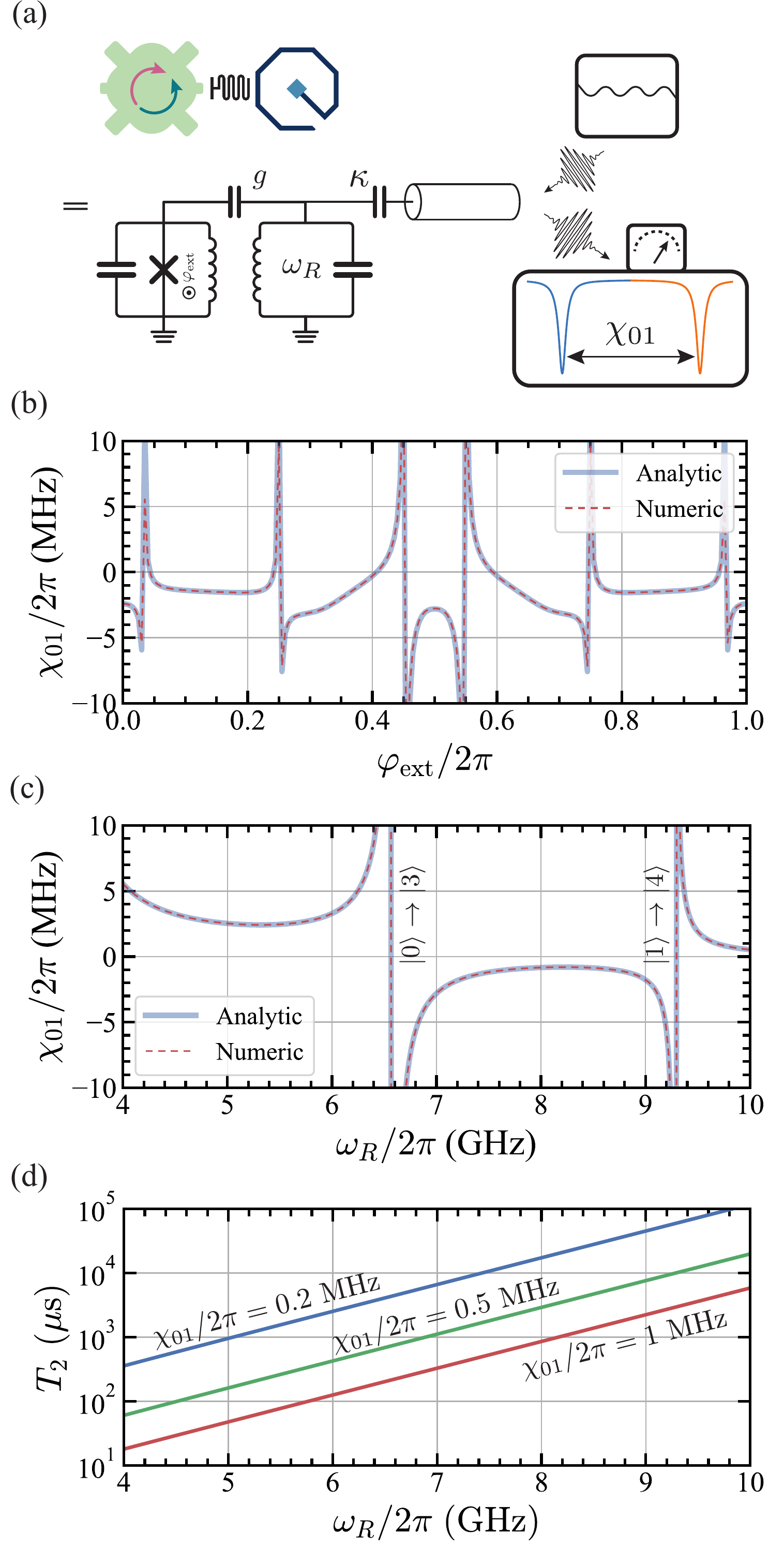}
        \caption{\label{fig3}Dispersive readout. \textbf{(a)} Schematic of a fluxonium qubit coupled to a readout resonator with coupling constant $g$. The resonator is then coupled to the readout bus with coupling rate $\kappa$, allowing measurement of its resonance $\omega_R$, which is shifted slightly from bare value, depending on the state of the qubit. No Purcell filter is needed. \textbf{(b)} Dispersive shift $\chi_{01}$ across one flux period for qubit parameters $\{E_J,E_C,E_L \} = \{4,1,1\}~\mathrm{GHz}$, resonator frequency $\omega_R/2\pi=7~\mathrm{GHz}$ and capacitive coupling constant $g/2\pi = 100~\mathrm{MHz}$. Numerical diagonalization results are the same for $p=1$ and $p=10$ photons. \textbf{(c)} Dispersive shift $\chi_{01}$ for qubit parameters $\{E_J,E_C,E_L \} = \{4,1,1\}~\mathrm{GHz}$ at flux bias $\varphi_\mathrm{ext}/2\pi=0.5$, corresponding to $\omega_{01}/2\pi \sim 0.58~\mathrm{GHz}$, coupling constant $g/2\pi = 100~\mathrm{MHz}$, and varying resonator frequency $\omega_R$. \textbf{(d)} Thermal photon dephasing time for the same qubit and resonator parameters in (c), with resonator temperature $T = 50~\mathrm{mK}$ and dispersive shift $\chi_{01}/2\pi=\{0.2,0.5,1\}~\mathrm{MHz}$, which can be changed by varying the geometric coupling constant $g$.}
    \end{figure}

    When an atom is off-resonantly coupled to a resonator, their dispersive interaction results in a resonator frequency dependent on the state of the qubit, allowing measurement of the latter by probing the former \cite{blais2007quantum,wallraff2005approaching}, as depicted in Fig.~\ref{fig3}(a). The generic Hamiltonian describing such a system can be written as \cite{zhu2013circuit}
    \begin{equation}
    \begin{split}
        \hat{H}_\mathrm{sys}/\hbar =  \sum_l \omega_l |l\rangle \langle l | &+ \sum_j \omega_j \hat{a}_j^\dagger \hat{a}_j \\
        &+ \sum_j \sum_{l,l'}g_{j;l,l'} |l\rangle\langle l'|(\hat{a}_j + \hat{a}_j^\dagger).
        \label{eqn:atom-oscillator}
        \end{split}
    \end{equation}
    Here, $\hat{a}^\dagger$ ($\hat{a}$) is the resonator's creation (annihilation) operator,  $l$ and $j$ respectively indicate the levels of the atom with frequency $\omega_l$ and resonator with frequency $\omega_j$. The coupling coefficient $g_{j;l,l'}$ depends on the geometric coupling constant $g_j$ between them and the corresponding matrix element of the qubit, $g_{j;l,l'}=g_j|\langle l|\hat{C}|l'\rangle|$, where $\hat{C}=\hat{n}$ for capacitive coupling and $\hat{C}=\hat{\varphi}$ for inductive coupling. 
    
    While selection rules in transmon limit the dispersive interaction to nearest levels \cite{koch2007charge}, the lack of such rules in fluxonium allows coupling between various qubit transitions and resonator modes. Thus, even for computational frequencies in the range below 1 GHz, the qubit's $|0\rangle \rightarrow |1\rangle$ transition still shifts the resonator via virtual transitions close to the latter's resonance. The dispersive regime is defined for this generic system following the condition $|\omega_{l,l'}-\omega_j|\gg g_{j;l,l'}\sqrt{\langle\hat{a}_j^\dagger\hat{a}_j\rangle+1}$.
    
    The generic Hamiltonian in Eq.~(\ref{eqn:atom-oscillator}) can be simplified to that describing a two-level system dispersively coupled to a single-mode harmonic oscillator with an effective dispersive shift $\chi_{01}$:
    \begin{equation}
    \label{eqn:dispersive_hamiltonian}
        \hat{H}_\mathrm{disp}/\hbar \approx \omega_{01} |1\rangle \langle 1| + (\omega_R + \chi_{01} |1\rangle \langle 1|)\hat{a}^\dagger \hat{a},
    \end{equation}
    where $\omega_{01}$ and $\omega_R\equiv \omega_{j=0}$ are respectively the qubit's $|0\rangle\rightarrow |1\rangle$ and resonator's resonance frequencies which are slightly Lamb-shifted \cite{schuster2005ac}. The structure of the charge matrix elements as shown in Fig.~\ref{fig1}(c), with $n_{01}$ small compared to others, is favorable for qubit-resonator capacitive coupling as the Purcell loss is further suppressed. In addition, capacitive coupling can be experimentally implemented in straightforward fashion, similar to the transmon case. The dispersive shift $\chi_{01}$ for qubit-resonator geometric coupling constant $g\equiv g_{j=0}$ can be derived using second-order perturbation theory as \cite{zhu2013circuit}
    \begin{equation}
        \chi_{01} = g^2 \left[\sum_{l\neq 0} |n_{0l}|^2 \frac{2\omega_{0l}}{\omega_{0l}^2 - \omega_R^2} - \sum_{l\neq 1}|n_{1l}|^2\frac{2\omega_{1l}}{\omega_{1l}^2 - \omega_R^2}  \right].
        \label{eqn:dispersive_shift_2nd}
    \end{equation}
    While the fourth-order corrections account for self-Kerr and cross-Kerr interactions, the additional corrections added to the dispersive shift given by Eq.~(\ref{eqn:dispersive_shift_2nd}) are negligible \cite{zhu2013circuit}. In the case where the perturbation approach breaks down, e.g. when the dispersive condition is not satisfied, numerical diagonalization of the coupled system \cite{smith2016quantization,smith2016fluxonium} may work best, and has been shown to consistently match experimentally extracted data up to hundreds of photons \cite{gusenkova2021quantum,takmakov2021minimizing}.
    
    For fluxonium parameters $\{E_J,E_C,E_L \} = \{4,1,1\}~\mathrm{GHz}$, we compute the dispersive shift of a cavity due to the qubit's $|0\rangle\rightarrow |1\rangle$ transition as a function of the external flux bias $\varphi_\mathrm{ext}$ using Eq.~(\ref{eqn:dispersive_shift_2nd}) with resonator frequency $\omega_R/2\pi = 7 ~\mathrm{GHz}$ and coupling constant $g/2\pi=100~\mathrm{MHz}$. For comparison, we numerically diagonalize a fluxonium-resonator system and extract the photon number dependent shift $\chi_{01}(p)=(\omega_{p+1,1} - \omega_{p,1}) - (\omega_{p+1,0} - \omega_{p,0})$, where $p$ is the number of resonator photons on average and the second subscript denotes the state of the qubit. As shown in Fig.~\ref{fig3}(b), the second-order approximation matches the numerical result perfectly for $p=[1-10]$ photons. Therefore, the simple relation given by Eq.~(\ref{eqn:dispersive_shift_2nd}) can be used to estimate the dispersive shift $\chi_{01}$ in most cases, avoiding the resource-intensive task of diagonalizing systems with large Hilbert space. Interestingly, although the qubit frequency goes from $\sim 5.5$ GHz at zero external flux to $\sim 0.58$ GHz at half-integer flux bias, the dispersive shift amplitude remains largely within a factor of two, with $\chi_{01}/2\pi\geq 1~\mathrm{MHz}$ across almost the entire flux period.
    
    To explore multiplexing capability of the readout, we simulate the dispersive shift $\chi_{01}$ with varying readout frequency as shown in Fig.~\ref{fig3}(c) for a qubit with parameters $\{E_J,E_C,E_L \} = \{4,1,1\}~\mathrm{GHz}$ biased at $\varphi_\mathrm{ext}/2\pi=0.5$, corresponding to qubit frequency $\omega_{01}/2\pi \sim 0.58~\mathrm{GHz}$. By inspecting Eq.~(\ref{eqn:dispersive_shift_2nd}) and the selection rules in Fig.~\ref{fig1}(c), we can expect a large shift when the cavity resonance is close to qubit transitions $|0\rangle \rightarrow |3\rangle$ or $|1\rangle \rightarrow |4\rangle$, and a plateau in between, which are confirmed by the simulation results. We obtain $|\chi_{01}|/2\pi \geq 1~\mathrm{MHz}$ for a nominal coupling constant $g/2\pi = 100~\mathrm{MHz}$ with the resonator frequency $\omega_R/2\pi$ ranging from $4$ to $10$ GHz, indicating a fast and flexible readout. In addition, the plateaus imply the same level of dispersive shift when there are fluctuations in qubit parameters. As a result, qubits with similar energy scales can be measured via individual resonators without further design constraints. We note that in addition to the small charge matrix element $n_{01}$, the qubit frequency $\omega_{01}/2\pi\sim 0.58~\mathrm{GHz}$ is much lower than the target resonator frequency, so relaxation through the resonator is strongly suppressed, eliminating the need for Purcell filters. 
    
    We must take into account the following considerations for the readout design. In case the resonator frequency is close to the qubit transitions, $|\omega_R - \omega_{\{03,14\}}| \sim g$, the dispersive shift can be large at the expense of short energy relaxation times associated with these transitions due to Purcell effect \cite{purcell1946resonance,houck2008controlling}. This may lead to dephasing of the computational states due to excitation to higher levels. For example, if the relaxation rate of the $|0\rangle \rightarrow |3\rangle$ transition is close to the resonator's linewidth, $\Gamma_{3\rightarrow 0}/2\pi \approx \kappa/2\pi = 2~\mathrm{MHz}$, and the qubit's effective temperature is $T_\mathrm{eff}^{0-3} = 50~\mathrm{mK}$ \cite{yan2018distinguishing}, the excitation rate would follow the principle of detailed balance, $\Gamma_{0\rightarrow 3}/2\pi\sim 10~\mathrm{kHz}$, corresponding to a dephasing time $T_\phi\approx 30~\mathrm{\mu s}$ for the $|0\rangle$ and $|1\rangle$ states. Therefore, it is best to avoid this scenario to protect the computational subspace from decoherence.
    
    
%
   There is inevitable thermal photon dephasing tradeoff \cite{zhang2017suppression,yan2018distinguishing} coming from the dispersive interaction with the resonator's fundamental mode, which we compute (see Appendix \ref{appendix:fluxonium}) and plot in Fig.~\ref{fig3}(d) for a resonator with varying resonance frequency $\omega_R$, using an effective temperature $T = 50~\mathrm{mK}$ \cite{yan2018distinguishing} and resonator linewidth $\kappa/2\pi = 2~\mathrm{MHz}$. In general, a resonator with frequency $\omega_R$ closer to the $|0\rangle\rightarrow |4\rangle$ transition has lower thermal photon, resulting in less dephasing. A compromise between readout signal which scales with $\chi_{01}$ \cite{blais2021circuit} and coherence time can also be made. 
   
   While virtual transitions $|0\rangle \rightarrow |3\rangle$ and $|1\rangle \rightarrow |4\rangle$ give rise to a finite dispersive shift, the matrix elements corresponding to higher-order transitions quickly vanish beyond 10 GHz, as shown in Fig.~\ref{fig1}(c). As a result, there is no coupling between the qubit and any higher modes of the readout resonator, and thus no additional dephasing machanisms associated with those modes. This justifies reducing the resonator model to a single mode in Eq.~(\ref{eqn:dispersive_hamiltonian}). 
   
   Although the dispersive shift should be reasonably small to reduce effect from thermal photon dephasing, as shown in Fig.~\ref{fig3}(d), a large number of resonator photons \cite{gusenkova2021quantum} can be used to improve the signal amplitude substantially, allowing high-fidelity readout for all fluxoniums around the bus. Moving forward, improvement of the cryogenic setup \cite{yeh2017microwave,yeh2019hot,yan2018distinguishing} to lower the resonator's effective temperature will reduce thermal photon population exponentially, enabling faster readout without compromising qubit coherence.
    
   The multiplexed readout platform can be arranged as shown in Fig.~\ref{fig0}, with four resonators spaced around a common bus. The dispersive shifts can be engineered to be within a plateau, so fluctuations in qubit frequencies would have negligible effect on readout performance. The resonator frequencies can be spaced sufficiently far apart to suppress off-resonant driving of untargeted resonators, which may lead to additional dephasing of qubits \cite{heinsoo2018rapid}. Since there is no need for Purcell filters \cite{reed2010fast,jeffrey2014fast,sete2015quantum}, the proposed platform's complexity and design constraints are reduced, leading to better scalability. The geometric coupling constant $g$ can be designed to target a dispersive shift $\chi_{01}/2\pi\sim 1~\mathrm{MHz}$, which facilitates fast readout while the coherence times can still be around 1 ms if the resonator frequency is chosen to be between $[8-9]~\mathrm{GHz}$, assuming a practical resonator temperature $T = 50~\mathrm{mK}$.
   
\section{\label{sec:single-qubit}Single-qubit gates}
    
Conventionally, single-qubit control in superconducting qubits is implemented by applying a radiation pulse on resonance with the transition frequency of the computational levels. Ideally, qubits consist of only two levels, so arbitrarily short square pulses are technically sufficient to make high-fidelity gates. In practice, quantum systems typically consist of many energy levels, $\hat{H}_q/\hbar=\sum_{j=1}^{d-1} \omega_j |j\rangle \langle j|$, and the transition frequencies differ by an amount $\alpha_j = (\omega_{j,j+1} - \omega_{j-1,j})/2\pi$, defined as the anharmonicity. In most cases, the lowest two levels of a quantum system are used as computational basis, and we consider the only relevant term, $\alpha \equiv \alpha_1$. To avoid inadvertently driving other levels, shaping the pulse to reduce its bandwidth is required \cite{bauer1984gaussian,steffen2003accurate}. 

\begin{figure}
    \includegraphics[width=0.47\textwidth]{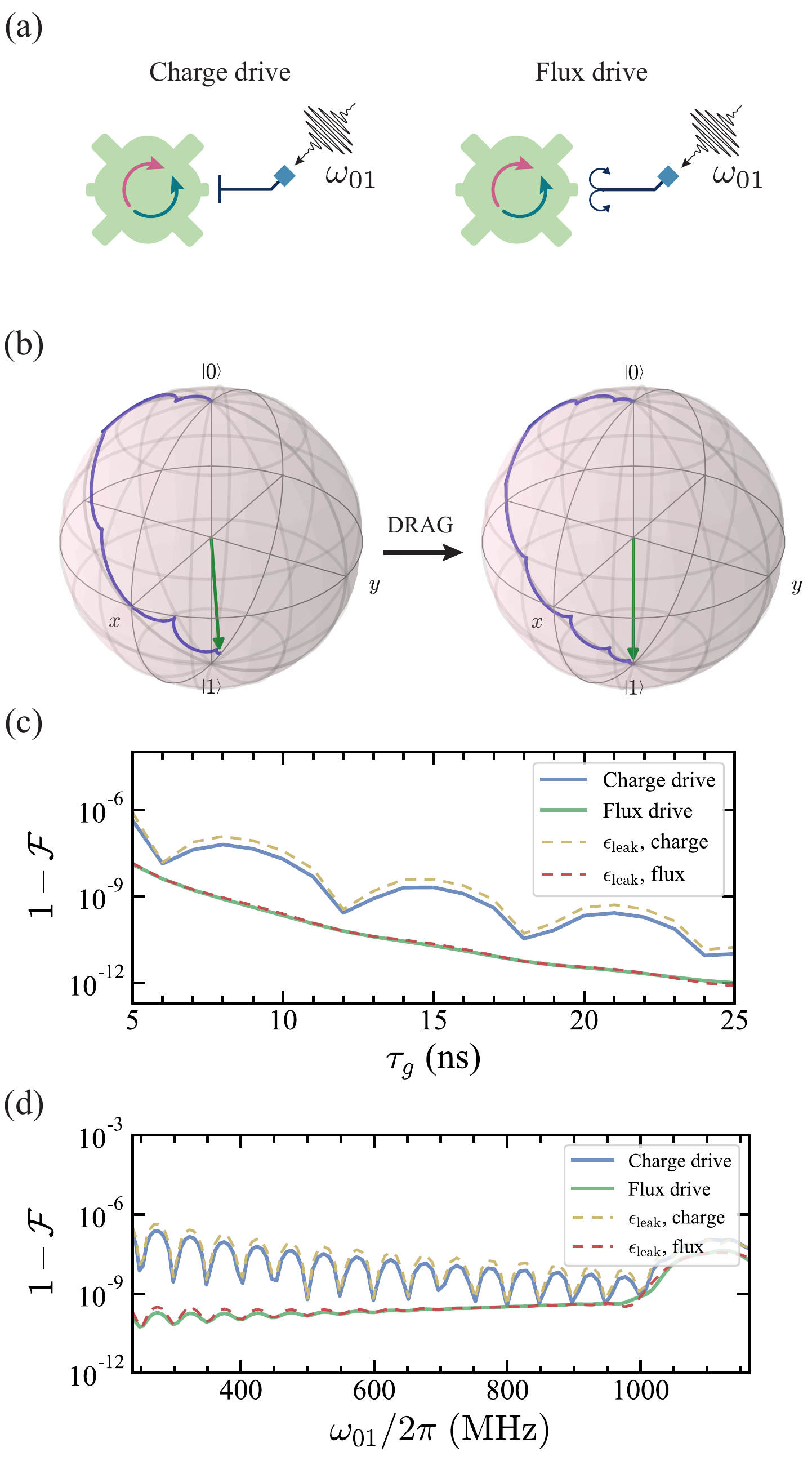}
    \caption{\label{fig2}Single-qubit gate. \textbf{(a)} Schematic showing microwave control of qubit via charge and flux drives. \textbf{(b)} Trajectory of qubit vector on the Bloch sphere before and after optimization using DRAG for flux driving with $\tau_g=10~\mathrm{ns}$. \textbf{(c)} Gate error and state leakage for varying gate time for both charge and flux coupling. \textbf{(d)} Gate error and leakage for varying qubit frequency, with gate time $\tau_g=10~\mathrm{ns}$. $\omega_{01}$ is changed by varying inductive energy $E_L$.}
\end{figure}

Intuitively, we expect errors in our operation when the pulse bandwidth approaches $\alpha$. In the extreme case of a harmonic oscillator, all the transition frequencies are equal, thus we inevitably excite other levels when we apply a resonant drive. In transmons, $\alpha=(\omega_{12} - \omega_{01})/2\pi\approx -E_C$, where $E_C$ is the charging energy, with typical value $E_C\sim 0.25~\mathrm{GHz}$. For fluxonium within our proposed parameter regime, $\alpha\sim 3~\mathrm{GHz}$, suggesting generally much lower single-qubit gate error using similar control techniques. However, the low qubit frequency implies that fast rotation does not follow the rotating wave approximation (RWA) \cite{krantz2019quantum}, so the dynamical evolution of the computational subspace may become intractable.
In this section, we explore the performance limit of single-fluxonium gates via both charge and flux microwave controls, which are respectively coupled capacitively and inductively to the qubit, as shown in the schematic in Fig.~\ref{fig2}(a). We focus on $X$- and $Y$-rotations induced by microwave drives, as $Z$ rotations are typically implemented virtually by the control software \cite{mckay2017efficient}.

First, we simulate single-qubit operations for the proposed fluxonium parameters $\{E_J, E_C, E_L\}=\{4,1,1\}~\mathrm{GHz}$, corresponding to qubit frequency $\omega_{01}/2\pi \sim 0.58~\mathrm{GHz}$ at external flux bias $\varphi_\mathrm{ext}/2\pi = 0.5$. Since the quantum dynamics in this low frequency subspace is susceptible to effects from the counter rotating terms, our approach is to numerically compute the qubit dynamics in the lab frame, then extract the unitary $\hat{U}_\mathrm{total}(t) = e^{i\phi} \hat{U}_\mathrm{comp}(t) \hat{U}_\mathrm{others}(t)$. Subsequently, we analyze the computational subspace evolution under $\hat{U}_\mathrm{comp}(t)$, disregarding the accumulated global phase $\phi$. The dynamics of higher states governed by $\hat{U}_\mathrm{others}(t)$ accounts for finite leakage out of the computational subspace. We present the results for $\{X,Y\}(\pi)$ gates here since they require higher pulse amplitudes for the same gate time compared to other rotation angles, and are thereby more susceptible to errors.

To this end, we couple a microwave pulse to the qubit's charge or phase degree of freedom. The pulse can be described by
\begin{equation}
    \mathcal{E} (t) = \mathcal{E}_I(t)\cos(\omega_d t) + \mathcal{E}_Q(t)\sin(\omega_d t),
\end{equation}
where $\mathcal{E}_I$ and $\mathcal{E}_Q$ are amplitudes of the in-phase and quadrature components of the pulse, and $\omega_d$ is the drive frequency. First, we apply an on-resonant in-phase pulse, $\omega_d = \omega_{01}$ and $\mathcal{E}_Q(t)=0$, with gate time $\tau_g =10~\mathrm{ns}$. The cosine envelope is chosen since its ramping is smooth and there is no need for truncation,
\begin{equation}
    \label{eqn:cosine_pulse}
    \mathcal{E}_I(t)= \frac{\epsilon_d}{2} \left[ 1-\cos(2\pi t/\tau_g) \right], 
\end{equation}
where the pulse amplitude $\epsilon_d$ is tuned up to implement a $\pi$ rotation, following the simple relation $\epsilon_d \eta_{01} \tau_g = 0.25$, with $\eta_{01}\equiv n_{01}$ for charge driving and $\eta_{01}\equiv \varphi_{01}$ for flux driving (see Appendix~\ref{appendix:sgate}). The qubit state's trajectory on the Bloch sphere is visualized in Fig.~\ref{fig2}(b, left), with its axis tilted due to the fast counter-rotating terms at $2\omega_d$. 

This can be corrected using derivative removal by adiabatic gate (DRAG) technique, which modifies the spectral profile of the pulse \cite{motzoi2009simple,gambetta2011analytic}. Specifically, the three control parameters we have, namely the envelopes $\mathcal{E}_I(t)$, $\mathcal{E}_Q(t)$, and the detuning $\delta = \omega_d - \omega_{01}$, can be tuned to effectively correct the tilt and improve gate fidelity, similar to the transmon case \cite{lucero2010reduced,chow2010optimized,chen2016measuring}. The first order correction is implemented by sending a quadrature pulse simultaneously with an amplitude proportional to the time-derivative of the in-phase pulse $\mathcal{E}_Q(t)=\lambda \dot{\mathcal{E}}_I(t)$, and adding a small detuning $\delta$. We find that this simple technique suppresses most of the gate error. The corrected trajectory for flux driving is shown in Fig.~\ref{fig2}(b, right). Optimization detail is discussed in Appendix~\ref{appendix:sgate}.

Next, we sweep the gate time $\tau_g$, optimizing the gate parameters for each instance, then compute the corresponding error $1-\mathcal{F}$, where the single-qubit gate fidelity $\mathcal{F}$ is defined as \cite{pedersen2007fidelity}
    \begin{equation}
   \label{eqn:single_q_fidelity}
       \mathcal{F} = \frac{1}{6} \left[\mathrm{Tr} (\hat{U}^\dagger \hat{U}) + |\mathrm{Tr}(\hat{U}^\dagger \hat{U}_\mathrm{ideal})|^2 \right],
    \end{equation}
with $\hat{U}\equiv \hat{U}_\mathrm{comp}$, which may not be unitary. The result in Fig.~\ref{fig2}(c) shows low errors for gate time as fast as $\tau_g=5~\mathrm{ns}$, and the error rate decreases exponentially for longer gate times, with error below $10^{-7}$ for $\tau_g=25~\mathrm{ns}$ in both charge and flux coupling cases. We attribute the residual errors to leakage, $\epsilon_\mathrm{leak}=1-(P_0+P_1)$, which are consistent with the simulation results.

By varying the inductive energy $E_L$ from 0.5 to 1.6 GHz with other energy parameters fixed, we can tune the qubit frequency $\omega_{01}/2\pi$ from $237~\mathrm{MHz}$ to $1163~\mathrm{MHz}$. We numerically simulate the gate error for a $\tau_g=10~\mathrm{ns}$ cosine pulse for both charge and flux coupling across this frequency range. The result in Fig.~\ref{fig2}(d) shows that the gate error, mainly coming from finite leakage, remains below $10^{-6}$. This promises high fidelity single-qubit operations in a large-scale device constructed from fluxoniums with approximately 1 GHz frequency bandwidth. More importantly, this implies that we can treat the circuit as a quasi-two-level system, especially for longer gate times.

Notably, due to the difference between flux and charge matrix elements discussed in Sec. ~\ref{sec:fluxonium}, charge coupling requires a stronger drive amplitude to implement the same rotation compared to flux coupling, so there is more leakage to higher levels for this case. Therefore, flux driving is better for operations involving computational states. Another advantage of flux coupling is that an RF flux drive can be combined with the DC flux bias using a diplexer, forming a single control line \cite{manenti2021full}, which is shown as the anchor in Fig.~\ref{fig0}. A symmetric flux line helps null parasitic capacitive coupling to the qubit, protecting it from energy decay \cite{hatridge2011dispersive}. A fast flux control line working at RF frequency has also been demonstrated to be compatible with high-coherence fluxonium qubit \cite{zhang2020fast}.

Typical microwave control of superconducting qubits involves IQ mixing of a low-frequency pulse with an RF carrier tone, which is susceptible to carrier leakage, imperfect sideband suppression, and pulse distortion due to non-ideal electronic performance. At the low qubit frequency regime we propose, the computational states can be controlled via microwave pulses synthesized directly from the arbitrary waveform generator (AWG), eliminating the need for IQ mixers and RF microwave sources. This would improve resource efficiency and reduce operational complexity significantly upon upgrading to large-scale devices.

Additionally, operating at low frequency gives us an advantage regarding microwave crosstalk, which is prevalent in superconducting qubit quantum processors. This spurious interaction between qubits and neighboring control lines may come from direct coupling, but is more likely due to radiation generated by an impedance mismatch at wire-bond pads, or coupling to common box-modes \cite{wenner2011wirebond,sheldon2017characterization,huang2021microwave,abrams2019methods}. Interestingly, crosstalk at the frequency range below 1 GHz is substantially smaller than at 5 GHz. This can be attributed to better impedance matching to the wire-bonds or disappearance of spurious modes at low frequencies \cite{wenner2011wirebond,huang2021microwave}. Such a feature will greatly benefit future quantum processors based on fluxonium qubits.

\section{\label{sec:multi-qubit}Multi-qubit coupling}

\begin{figure*}[t!]
        \includegraphics[width=0.75\textwidth]{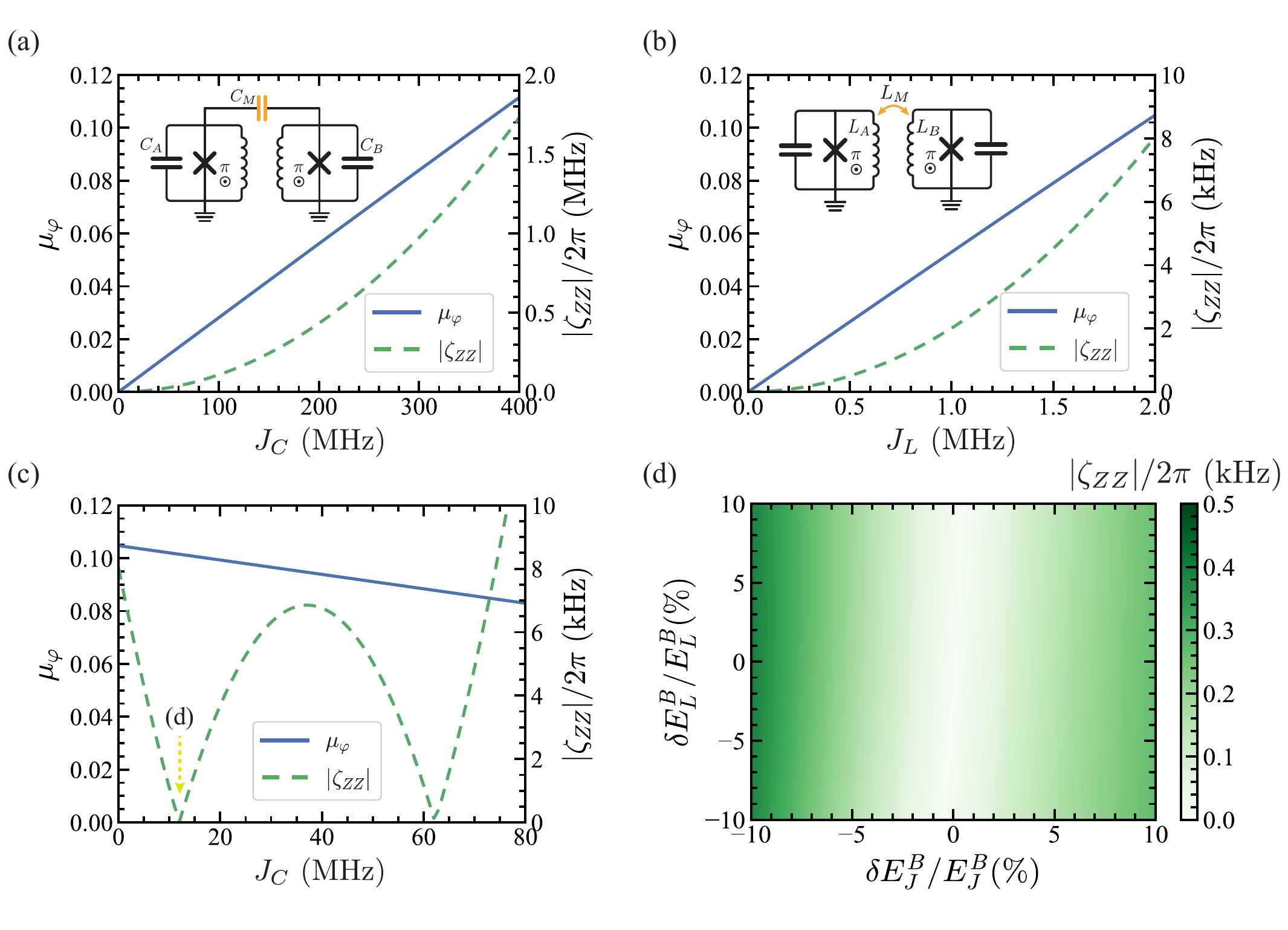}
        \caption{\label{fig5}Multi-fluxonium coupling for qubit parameters listed in Table~\ref{tab:coupled_qubit_params}. Normalized matrix element $\mu_\varphi$ and static longitudinal coupling rate $\zeta_{ZZ}$ for varying \textbf{(a)} capacitive coupling $ \hat{H}_\mathrm{coupl}=J_C\hat{n}_A\hat{n}_B$, \textbf{(b)} inductive coupling $ \hat{H}_\mathrm{coupl}=- J_L \hat{\varphi}_A\hat{\varphi}_B$, and \textbf{(c)} multi-path coupling $ \hat{H}_\mathrm{coupl}=J_C\hat{n}_A\hat{n}_B - J_L \hat{\varphi}_A\hat{\varphi}_B$ with $J_L=2~\mathrm{MHz}$. \textbf{(d)} Variation of static longitudinal coupling rate $\zeta_{ZZ}$ due to fluctuations in qubit B's parameters for coupling coefficients $\{J_L,J_C\}=\{2,11.5\}~\mathrm{MHz}$.}
    \end{figure*}

In this section, we explore different coupling scenarios and discuss their properties. We show that a multi-path coupling technique can be used to statically suppress the spurious $ZZ$ rate across the whole lattice while allowing the computational states to have sufficiently strong interaction for fast entangling gate operations.

A quantum system consisting of two directly coupled fluxonium qubits can be described by the Hamiltonian \cite{nesterov2018microwave}
\begin{equation}
    \hat{H} =  \hat{H}_A +  \hat{H}_B +  \hat{H}_\mathrm{coupl},
    \label{eqn:dcoupled_fluxonium_hamiltonian}
\end{equation}
where $ \hat{H}_{A,B}$ is the bare (uncoupled) Hamiltonian given by Eq.~(\ref{eqn:fluxonium_hamiltonian}), and the coupling term can be written as $ \hat{H}_\mathrm{coupl}/h = J_C \hat{n}_A \hat{n}_B$ for capacitive coupling and $ \hat{H}_\mathrm{coupl}/h = -J_L \hat{\varphi}_A \hat{\varphi}_B$ for inductive coupling. Here, $\hat{n}_{A,B}$ and $\hat{\varphi}_{A,B}$ are respectively the charge and phase degrees of freedom of qubit $A$ and qubit $B$. The coupling coefficients are proportional to the mutual circuit element, $J_C = 4e^2C_M/(C_AC_B)$ \cite{devoret1995quantum,vool2017introduction} and $J_L=(\hbar/2e)^2L_M/(L_AL_B)$ \cite{smith2016fluxonium,smith2016quantization} in the weak coupling regime where $C_M\ll C_A,C_B$ and $L_M\ll L_A,L_B$. 
    
    \begin{table}[b]
    \caption{\label{tab:coupled_qubit_params}Coupled fluxonium parameters.}
        \begin{tabular}{||c||c|c|c|c||} 
         \hline
         \rule{0pt}{2ex}Qubit & $E_J$ (GHz)  & $E_C$  (GHz) & $E_L$ (GHz) & $\omega_{01}/2\pi$ (MHz)\\ [0.5ex] 
         \hline\hline
         \rule{0pt}{2ex}A & 4.0 & 1.0 & 0.9 & 499 \\ 
         \hline
         \rule{0pt}{2ex}B & 4.0 & 1.0 & 1.0 & 581\\ [0ex] 
         \hline 
        \end{tabular}
    \end{table}
    
    We use the notation $|k_A l_B\rangle$ to indicate states of interacting systems in the rest of this work. The bare states of uncoupled qubits are denoted as $|k_A l_B\rangle_0$. Here, we focus on using multi-qubit gates based on the interaction between computational states, which causes mixing between $|01\rangle$ and $|10\rangle$. To characterize the level of their hybridization, we compute the normalized cross matrix element defined as 
    \begin{equation}
    \label{eqn:normalized_matrix_element}
        \mu_\varphi = \frac{|\langle 00 | \hat{\varphi}_A\otimes\hat{I}_B | 01 \rangle|}{|\langle 00 | \hat{\varphi}_A\otimes\hat{I}_B | 10 \rangle|}.
    \end{equation}
    This quantity is zero for uncoupled system since $_0\langle 00 | \hat{\varphi}_A\otimes\hat{I}_B | 01 \rangle_0 = 0$, and becomes finite due to dressing between the computational states. The normalization form defined by Eq.~(\ref{eqn:normalized_matrix_element}) is chosen such that the corresponding matrix element amplitude is taken into account. For example, the normalized cross matrix element for charge operator $\hat{n}_A\otimes \hat{I}_B$ can be used to characterize the mixing as well. 
    
    As the amplitude of $\mu_\varphi$ does not carry practical information, we can compare it with an equivalent quantity in a system of coupled spins described by the Hamiltonian
    \begin{equation}
    \label{eqn:coupled_spins}
         \hat{H}_\mathrm{ss}/\hbar=\frac{1}{2}\omega_A \hat{Z}_A + \frac{1}{2}\omega_B \hat{Z}_B + 2\pi J_\mathrm{eff}\hat{X}_A\hat{X}_B.
    \end{equation}
     For $\omega_A/2\pi=499~\mathrm{MHz}$, $\omega_B/2\pi=581~\mathrm{MHz}$, and $J_\mathrm{eff}=10$ MHz, this system has a normalized cross matrix element $\mu_X = 0.11$, where
     \begin{equation}
         \mu_X = \frac{|\langle 00 | \hat{X}_A\otimes\hat{I}_B |  01 \rangle|}{|\langle  00 | \hat{X}_A\otimes\hat{I}_B |  10 \rangle|}.
     \end{equation}
     Fast microwave-entangling gates in coupled transmons system with similar $J_\mathrm{eff}$ have been demonstrated \cite{hashim2020randomized,mitchell2021hardware}, so we shall use this $\mu_X$ value as a reference. More detail on the mapping between coupled spins model and interacting fluxoniums system is discussed in Appendix \ref{appendix:coupling}.
    
    After diagonalizing the Hamiltonian in Eq.~(\ref{eqn:dcoupled_fluxonium_hamiltonian}) with parameters listed in Table \ref{tab:coupled_qubit_params} for varying charge and flux coupling constants, we extract $\mu_\varphi$ together with the static longitudinal coupling rate defined as
    \begin{equation}
        \zeta_{ZZ} = \omega_{|00\rangle}+\omega_{|11\rangle} - \omega_{|10\rangle} - \omega_{|01\rangle},
    \end{equation}
    where $\omega_{|k_Al_B\rangle}$ is the eigen-energy of the two-qubit state $|k_Al_B\rangle$. This static $ZZ$ can be viewed as an always-on entangling operation, leading to errors in both local and non-local operations in a large-scale processor \cite{sundaresan2020reducing}. Our primary goal is thus to enable a sufficiently large exchange interaction characterized by $\mu_\varphi$, and at the same time to suppress $\zeta_{ZZ}$.
    
    As shown in Fig.~\ref{fig5}(a), capacitive coupling results in a large $ZZ$ rate to achieve a level of dressing corresponding to $\mu_\varphi\sim 0.11$. This can be understood as follows. Since the longitudinal coupling is a dispersive effect, interactions between higher levels play an important role. Meanwhile, the charge matrix elements follow $n_{01}< n_\mathrm{others}$ (Fig.~\ref{fig1}(c)), resulting in strong mixing between non-computational states compared to computational states. This leads to large $\zeta_{ZZ}$ in order to reach a desired $\mu_\varphi$. On the other hand, as shown in Fig.~\ref{fig5}(b), inductive coupling gives a rather small $\zeta_{ZZ}$ to reach the same $\mu_\varphi$, because $\varphi_{01} > \varphi_\mathrm{others}$. Therefore, capacitive coupling does not provide a sufficient exchange interaction between computational states but is better for tuning $ZZ$, and vice versa for inductive coupling.
    
     Based on these selection rules, we propose a multi-path coupling approach that includes an inductive coupling term $-J_L\hat{\varphi}_A\hat{\varphi}_B$ that enables the exchange interaction in the computational states, and a small capacitive coupling term $J_C\hat{n}_A\hat{n}_B$ that can be used to suppress the residual static $ZZ$. As shown in Fig.~\ref{fig5}(c), for qubits with parameters listed in Table~\ref{tab:coupled_qubit_params} and inductive coupling constant $J_L=2~\mathrm{MHz}$ (corresponding to $L_M=91~\mathrm{pH}$), a small capacitive coupling with coefficient $J_C = 11.5~\mathrm{MHz}$ (corresponding to $C_M=26~\mathrm{aF}$) is needed to make $|\zeta_{ZZ}|=0$.
    
    Figure $\ref{fig5}$(d) further shows that even when the fluxonium parameters fluctuate within $10\%$, the static $ZZ$ rate for a pair of qubits coupled in multi-path fashion would remain below 1 kHz, hence the scheme is resilient against parameter fluctuations. Interestingly, $\zeta_{ZZ}$ stays close to zero across a wide range of $E_L^B$, so the same coupling design parameters can be used for many qubit pairs if $\omega_{01}$ is tuned by varying $E_L$, promising another advantage upon scaling up.
    
    The primary challenge in scaling up this multi-path coupling scheme will be to combine the large inductive coupling with $L_M\sim 100~\mathrm{pH}$ and the small capacitive coupling with $C_M\sim 25~\mathrm{aF}$. For the inductive part, the most straightforward approach is to design neighboring qubits to share sections of their superinductors \cite{kou2017fluxonium}. Since the qubits must be placed next to each other in this case, classical microwave control crosstalk can become significant. Future development of a coupling element \cite{chen2014fabrication,yan2018tunable,leroux2021superconducting} or an inductive bus analogous to the resonator bus in transmon architectures \cite{blais2007quantum,majer2007coupling} would be ideal. For the capacitive part, microwave design following standard transmon techniques will be sufficient to realize the required coupling. Since the proposed charging energy $E_C$ parameter corresponds to a small antenna, and the ground plane will screen a large portion of qubit-qubit cross capacitances, it is straightforward to engineer the desired mutual capacitance.

\section{\label{sec:two_qubit_gates}Multi-qubit gates}
In this section, we investigate the performance of two types of microwave-activated entangling operations in the low frequency regimes for $J_L= 2~\mathrm{MHz}$ and $J_C = 11.5~\mathrm{MHz}$ to cancel the static $ZZ$ rate, corresponding to $J_\mathrm{eff}\approx 11~\mathrm{MHz}$ in Eq.~(\ref{eqn:coupled_spins}). Having shown that leakage outside the computational subspace is negligible, $\epsilon_\mathrm{leak}<10^{-6}$ for sufficiently long gate time in Sec.~\ref{sec:single-qubit}, and that the static $ZZ$ can be cancelled using multi-path coupling in Sec.~\ref{sec:multi-qubit}, we may reduce the complex multi-fluxonium system to a simpler model describing two coupled spins. We discuss this mapping in more detail in Appendix~\ref{appendix:coupling}, and proceed to use the practical formalism describing two-level systems here, denoting $\omega_A\equiv \omega_{01}^A$ and $\omega_B\equiv \omega_{01}^B$.

The entangling gates we study are the cross-resonance (CR) CNOT \cite{de2010selective,rigetti2010fully,chow2011simple} and differential AC-Stark CZ \cite{mitchell2021hardware,wei2021quantum} gates, which can be used to implement quantum error correction codes \cite{terhal2015quantum,campbell2017roads}. These microwave gates only involve the high-coherence computational states, so the errors resulting from decoherence are small. In addition, they can be readily implemented using the same single-qubit control components on the chip, in this case the RF flux lines, reducing the design and fabrication constraints overhead upon scaling up. We note that, while not explored here, other entangling gates such as the AC-Stark shift \cite{blais2007quantum,majer2007coupling} and the two-photon \cite{poletto2012entanglement,nesterov2021proposal} $\mathrm{SWAP}$-like gates are also compatible with our proposed architecture.

    \subsection{Cross-resonance controlled-NOT gate}
    
    \begin{figure*}
        \includegraphics[width=\textwidth]{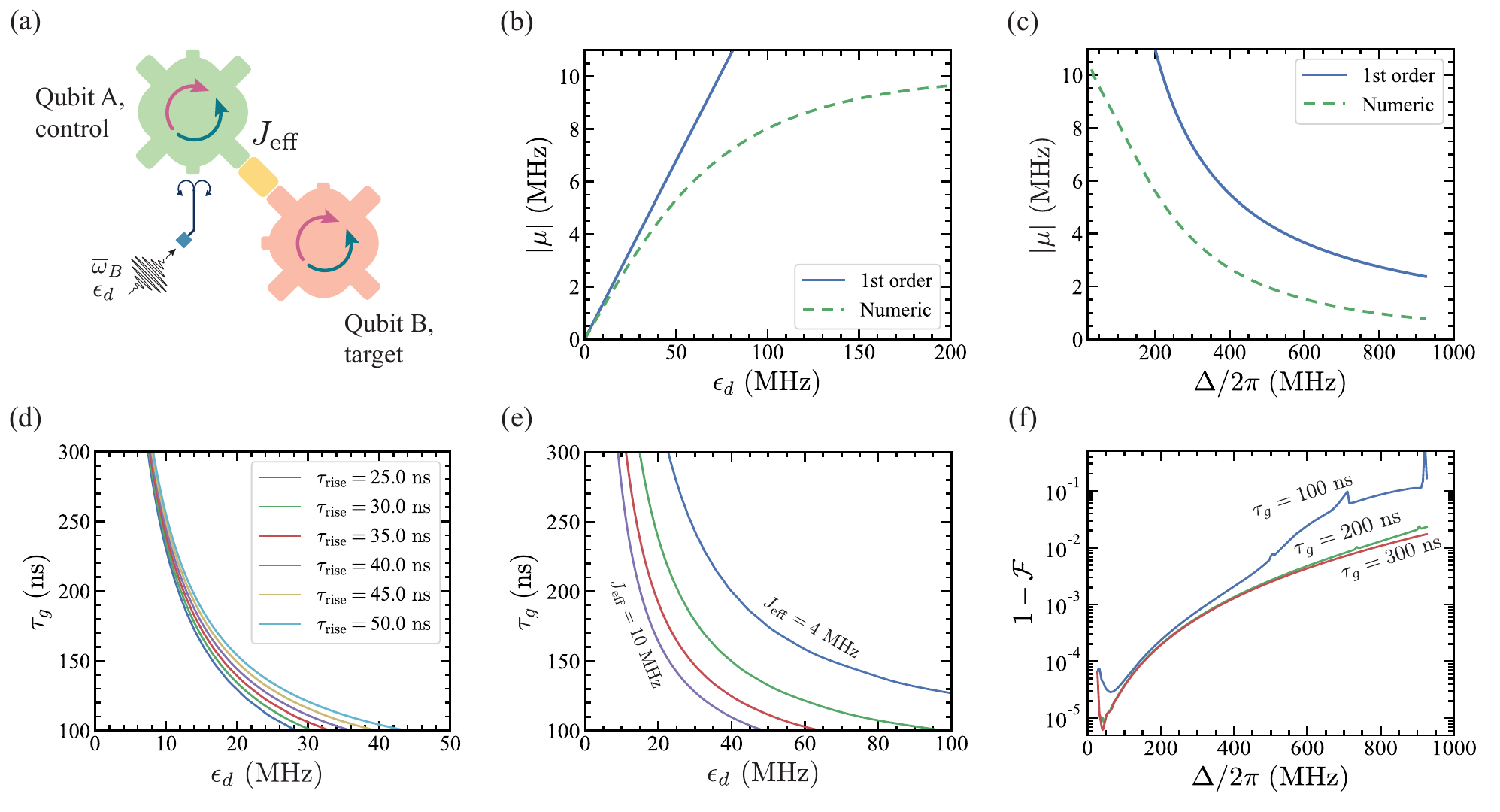}
        \caption{\label{fig6}Cross-resonance gate simulation. Except for plot (e), the coupling coefficients are $J_L=2~\mathrm{MHz}$ and $J_C=11.5~\mathrm{MHz}$, corresponding to an effective exchange interaction rate $J_\mathrm{eff} \sim 11~\mathrm{MHz}$. Qubit parameters are listed in Table \ref{tab:coupled_qubit_params} for plots (b), (d), and (e). Qubit B's inductive energy $E_{L,B}$ is changed to vary the qubit-qubit detuning $\Delta$ as listed in Table~\ref{tab:coupled_qubit_sweep} for plots (c) and (f). \textbf{(a)} Gate scheme: a microwave pulse applied to qubit A at the dressed frequency $\overline{\omega}_B$ of qubit B will induce an effective $ZX$ interaction which entangles the qubits. \textbf{(b)} Effective $ZX$ rate $\mu$ with varying drive amplitude $\epsilon_d$. \textbf{(c)} Effective $ZX$ rate $\mu$ with varying qubit-qubit detuning $\Delta=\omega_{B}-\omega_{A}$ for drive amplitude $\epsilon_d=200~\mathrm{MHz}$. \textbf{(d)} CNOT gate time $\tau_g$ for  varying drive amplitude $\epsilon_d$, with
        ramping time $\tau_\mathrm{ramp}=\{25,30,35,40,45,50\}~\mathrm{ns}$. \textbf{(e)} CNOT gate time $\tau_g$ for varying drive amplitude $\epsilon_d$ with
        different effective exchange interaction strength $J_\mathrm{eff}\sim \{4,6,8,10\}~\mathrm{MHz}$ and gate ramping time $\tau_\mathrm{ramp}=50~\mathrm{ns}$. \textbf{(f)} Gate error for varying qubit-qubit detuning $\Delta$ with pulse ramping time $\tau_\mathrm{ramp}=50~\mathrm{ns}$, and total gate time $\tau_g=\{100,200,300\}~\mathrm{ns}$.}
    \end{figure*}
    
    Among various microwave-activated entangling operations, the CR gate has been the most popular for fixed-frequency superconducting qubits. It was first proposed considering a model of two coupled two-level systems \cite{paraoanu2006microwave,rigetti2010fully}, and subsequently implemented using flux qubits \cite{de2010selective,chow2011simple}, followed by its application on transmons \cite{chow2012universal} with a gate fidelity of 95\%. Substantial improvements to the gate were made using echo sequence \cite{corcoles2013process} and simultaneous cancellation tone \cite{sheldon2016procedure}, resulting in a gate fidelity of $\mathcal{F}=99.1\%$. Recent tune-up procedure involving rotary echo sequence has been used to demonstrate high fidelity uniformly in a multi-qubit quantum processor \cite{sundaresan2020reducing}. On the theory side, after initial study of the gate \cite{paraoanu2006microwave,rigetti2010fully,de2012selective}, the need to understand and optimize the two-qubit CR interaction in coupled-transmons systems more thoroughly has led researchers to analyze it more rigorously \cite{tripathi2019operation,magesan2020effective,malekakhlagh2020first}, with optimal pulse shaping promising to improve the gate even further \cite{kirchhoff2018optimized}. Recently, a single-step high-fidelity three-qubit iToffoli entangling gate based on CR interactions has been experimentally demonstrated, providing a more expressive gate set and possible reduction in circuit depth for NISQ algorithms \cite{kim2021high}.
    
    Figure \ref{fig6}(a) depicts the implementation of the gate. Two coupled spins described by the Hamiltonian in Eq.~(\ref{eqn:coupled_spins}) would have transition frequencies shifted due to the dressing of the states, $\overline{\omega}_A=\omega_A - 2\pi J_\mathrm{eff}/\Delta$,  $\overline{\omega}_B=\omega_B + 2\pi J_\mathrm{eff}/\Delta$, where $\Delta=\omega_B - \omega_A$ is the frequency detuning between the qubits. If the coupling $J_\mathrm{eff}$ is finite, a microwave tone applied to qubit A at the frequency of qubit B will excite qubit B, and its evolution pattern will depend on the state of qubit A. Thus, the qubits can become entangled. Qubit A and qubit B are usually referred to in the literature as the \textit{control} and \textit{target} qubits, respectively \cite{rigetti2010fully}.
    
    We can approximate the entangling rate at the lowest order as follows. The microwave drive with amplitude $\epsilon_d$ induces the CR effect that manifests as \cite{tripathi2019operation}
    \begin{equation}
        \label{eqn:cr_eff1}
        \hat{H}_\mathrm{CR,eff}/h= (\epsilon_0{|01\rangle }\, {\langle 00|} + \epsilon_1{|11\rangle}\, { \langle 10|}) + \mathrm{h.c.},
    \end{equation}
    where $\epsilon_0$ and $\epsilon_1$ can be viewed as the effective drive amplitudes on the target qubit when the control qubit is in states $|0\rangle$ and $|1\rangle$, respectively. The Hamiltonian in Eq.~(\ref{eqn:cr_eff1}) can be rewritten using operator format to highlight the important $ZX$ term \cite{rigetti2010fully,chow2011simple},
    \begin{equation}
        \label{eqn:cr_eff2}
        \hat{H}_\mathrm{CR,eff}/h= m\hat{I}\hat{X} + \mu \hat{Z}\hat{X},
    \end{equation}
    where the $IX$ coefficient is $m=(\epsilon_0 + \epsilon_1)/2$ in the absence of classical microwave crosstalk, and the $ZX$ amplitude is $\mu = (\epsilon_0 - \epsilon_1)/2$. The effective drive coefficients in Eq.~(\ref{eqn:cr_eff1}) can be approximated at the first order as $\epsilon_0\approx-(2\pi J_\mathrm{eff}/\Delta)\epsilon_d$, $\epsilon_1\approx(2\pi J_\mathrm{eff}/\Delta)\epsilon_d$, which gives $m\approx 0$ and $\mu \approx (2\pi J_\mathrm{eff}/\Delta)\epsilon_d$ \cite{tripathi2019operation}. We note that in practice, there is a finite $IX$ amplitude from microwave crosstalk or when there is participation from other qubit levels, resulting in $m>0$. The $ZI$ term due to the control qubit's AC-Stark shift is also omitted in Eq.~(\ref{eqn:cr_eff2}). In principle, these single-qubit operators commute with the desired $ZX$ term, and hence do not degrade the gate fidelity.  
    
    To estimate the gate rate, which dictates how fast the qubits can be entangled, we compute the $ZX$ amplitude $\mu$ for a pair of coupled fluxoniums with parameters listed in Table \ref{tab:coupled_qubit_params}, corresponding to bare qubit frequencies $\omega_A/2\pi=499~\mathrm{MHz}$, $\omega_B/2\pi=581~\mathrm{MHz}$. As mentioned above, we use coupling coefficients $J_L = 2~\mathrm{MHz}$ and $J_C = 11.5~\mathrm{MHz}$, corresponding to an effective spin-spin exchange interaction rate $J_\mathrm{eff} \approx 11~\mathrm{MHz}$. In addition, we numerically compute the gate rate by applying a continuous microwave tone to qubit A at frequency $\overline{\omega}_B$ and extracting the oscillation frequency of qubit B \cite{sheldon2016procedure}. The numerical procedure involves up to five levels in each qubit which is sufficient to describe the essential dynamics and possible errors.
    
    As shown in Fig.~\ref{fig6}(b), the lowest order estimation $\mu\approx (2\pi J_\mathrm{eff}/\Delta) \epsilon_d$ agrees well with the numerically obtained result in the small drive amplitude region, where a perturbation approach is supposed to work well. However, at large drive amplitude, the rate approaches a plateau, indicating a saturated gate rate. Therefore, the gate cannot be made arbitrarily short by simply increasing the drive amplitude.
    
    Next, we explore how the gate rate varies with different qubit-qubit detunings, which is important for large-scale devices. To this end, we fix qubit A's parameters, with its inductive energy $E_{L,A}=0.5~\mathrm{GHz}$, and sweep qubit B's inductive energy $E_{L,B} = [0.55-1.6]~\mathrm{GHz}$, corresponding to qubit frequencies $\omega_A/2\pi = 237~\mathrm{MHz}$ and $\omega_B/2\pi = [264-1163]~\mathrm{MHz}$, as listed in Table.~\ref{tab:coupled_qubit_sweep}. We repeat the simulation with these parameters, corresponding to qubit-qubit detuning $\Delta/2\pi = [27-926]~\mathrm{MHz}$, using a large drive amplitude $\epsilon_d = 200~\mathrm{MHz}$. Fig.~\ref{fig6}(c) shows the difference between the first-order-approximation and the numerical results, with the latter showing a large but finite gate rate $|\mu| > 5~\mathrm{MHz}$ at small detuning, and a small gate rate $|\mu| <1~\mathrm{MHz}$ as the detuning approaches 1 GHz.
    
    \begin{table}[b]
    \caption{\label{tab:coupled_qubit_sweep}Coupled fluxonium parameters for simulation involving sweeping of the qubit-qubit detuning $\Delta=\omega_B-\omega_A$.}
        \begin{tabular}{||c||c|c|c|c||} 
         \hline
         Qubit & $E_J$ (GHz)  & $E_C$  (GHz) & $E_L$ (GHz) & $\omega_{01}/2\pi$ (MHz)\\ [0.5ex] 
         \hline\hline
         \rule{0pt}{2ex}A & 4.0 & 1.0 & 0.5 & 237 \\ 
         \hline
         \rule{0pt}{2ex}B & 4.0 & 1.0 & [0.55-1.6] & [264-1163]\\ [0ex] 
         \hline 
        \end{tabular}
    \end{table}
    
    Having established how the drive amplitude $\epsilon_d$ and qubit-qubit detuning $\Delta$ determine the gate rate, we proceed to calibrate the correct pulse and simulate the performance of the gate. The CR operation can be used to implement a CNOT unitary,
    \begin{equation}
    \label{eqn:cnot}
        \hat{U}_\mathrm{CNOT} = \mathrm{exp}\left[\frac{i\pi}{4}(\hat{Z}\hat{I}+\hat{I}\hat{X}-\hat{Z}\hat{X}-\hat{I}\hat{I})\right].
    \end{equation}
    The $ZI$ and $IX$ terms commute with $ZX$ and can be implemented using single-qubit gates, while the $II$ term simply introduces a global phase. Hence, to realize a CNOT gate, we have to apply the CR pulse with time-dependent amplitude $\mathcal{E}(t)$ for a time $\tau_g$ satisfying the condition 
    \begin{equation}
        \int_{0}^{\tau_g} \mu (t) dt = \frac{\pi}{2},
    \end{equation}
    then add single-qubit gates which have low errors as discussed in Sec.~\ref{sec:single-qubit}.
    
    In practice, we tune up the gate numerically by initializing the qubits in the ground states $|00\rangle$, applying the drive with amplitude $\epsilon_d$ as a parameter, and optimizing the conditionality $R$ \cite{sheldon2016procedure,mitchell2021hardware}, given as 
   \begin{equation}
   \label{eqn:conditional}
   \begin{split}
       R = 
     \frac{1}{2} [ &(\langle \hat{X}\rangle_0 - \langle \hat{X}\rangle_1)^2 \\
     + &(\langle \hat{Y}\rangle_0 - \langle \hat{Y}\rangle_1)^2 + (\langle \hat{Z}\rangle_0 - \langle \hat{Z}\rangle_1)^2]^{\frac{1}{2}},
    \end{split}
   \end{equation}
   where, for example, $\langle \hat{X}\rangle_0$ is the expectation value of the target qubit's operator $\hat{X}$ when the control qubit is in state $|0\rangle$. $R$ can also be viewed as an entanglement metric, with $R=0$ for unentangled states and $R=1$ for maximally entangled states. We can find the correct gate parameters by varying the gate time and amplitude to minimize $| 1-R |$. To avoid microwave leakage and at the same time ensure sufficient drive amplitude, we use a flat-top pulse envelope with cosine-ramping at both ends,
        \[ 
        \label{eqn:pulse_shape}
            \mathcal{E}(t) =
            \begin{cases}
                 \frac{\epsilon_d}{2}\left[1-\cos\frac{\pi t}{\tau_\mathrm{ramp}}\right] &, 0 \leq t \leq \tau_\mathrm{ramp} \\
                \epsilon_d &, \tau_\mathrm{ramp} \leq t \leq \tau_g - \tau_\mathrm{ramp} \\
                \frac{\epsilon_d}{2} \left[1-\cos\frac{\pi (\tau_g-t)}{\tau_\mathrm{ramp}}\right] &, \tau_g - \tau_\mathrm{ramp} \leq t \leq \tau_g.
            \end{cases}
        \]
        
   To tune up the correct parameters, we first compute the corresponding gate time $\tau_g$ for varying drive amplitude $\epsilon_d$, with different ramping time $\tau_g$. The results in Fig.~\ref{fig6}(d) show that shorter gate time and longer ramping time require stronger drive amplitudes, as expected. We note that in principle, the effective coupling can also be engineered to allow faster gate time for the same drive amplitude, as simulated in Fig.~\ref{fig6}(e).
   
   Then, we simulate the performance of the gate by computing its fidelity with varying qubit-qubit detuning as listed in Table~\ref{tab:coupled_qubit_sweep}, using pulse ramping time $\tau_\mathrm{ramp} = 50~\mathrm{ns}$ and pulse duration $\tau_g=\{100,200,300\}~\mathrm{ns}$. We optimize the pulse amplitude at each parameter point using Nelder-Mead method to estimate the best gate fidelity. Figure~\ref{fig6}(f) shows the gate error $1-\mathcal{F}$, where the average gate fidelity $\mathcal{F}$ is defined for two-qubit gate as \cite{pedersen2007fidelity}
   \begin{equation}
   \label{eqn:two_q_fidelity}
       \mathcal{F} = \frac{1}{20} \left[\mathrm{Tr} (\hat{U}^\dagger \hat{U}) + |\mathrm{Tr}(\hat{U}^\dagger \hat{U}_\mathrm{ideal})|^2 \right].
   \end{equation}
    
    This shows that the fidelity generally gets worse for large detuning, with gate error ranging from around $10^{-5}$ at $\Delta/2\pi=30~\mathrm{MHz}$ to $10^{-2}$ at $\Delta/2\pi=900~\mathrm{MHz}$. Since the gate rate $\mu$ saturates steadily for large detuning $\Delta$ and strong drive amplitude $\epsilon_d$, we can anticipate the gate to become worse when an appropriate $\epsilon_d$ cannot be found. Notably, slightly longer gate time does not improve the fidelity substantially, so the best approach to achieve better gate fidelity at large detuning would be to increase the coupling rate and induce a higher gate rate for the same drive amplitude (cf. Fig.~\ref{fig6}(e)). Because the static $ZZ$ rate is cancelled via the multi-path coupling, stronger coupling should not introduce any adverse effect.
    
    Since the drive frequency in the CR gate scheme is fixed to the target qubit, there can be stringent requirements on adjacent qubits' parameters to avoid frequency collisions. In qubit systems with low anharmonicity, in addition to avoiding the collision of $|0\rangle\rightarrow|1\rangle$ transitions between qubits, we also have to take into account quantum dynamics involving $|1\rangle\rightarrow|2\rangle$ transitions and multi-photon processes \cite{malekakhlagh2020first}. Our proposed architecture leverages the high anharmonicity of fluxonium to alleviate these requirements. This allows the gate to be reasonably fast with fidelity $\mathcal{F}\sim 0.99$ at large detuning.
    
\subsection{Differential AC-Stark controlled-Z gate}
    \begin{figure*}
        \includegraphics[width=\textwidth]{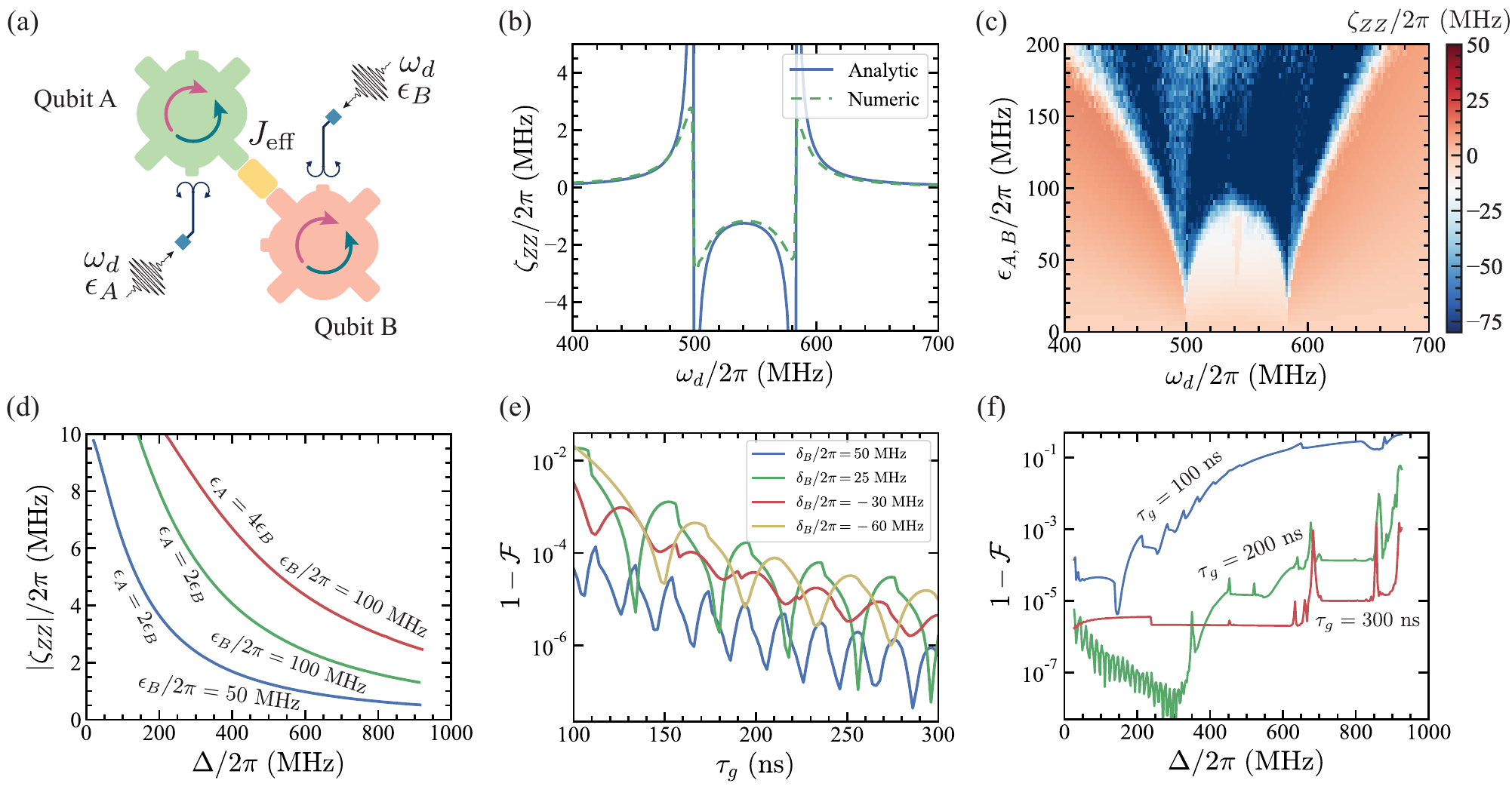}
        \caption{\label{fig7}Differential AC-Stark CZ gate simulation. The coupling coefficients are $\{J_L,J_C\} = \{2,11.5\}~\mathrm{MHz}$, corresponding to an effective exchange interaction rate $J_\mathrm{eff}\sim 11~\mathrm{MHz}$. Qubit parameters with fixed detuning are listed in Table~\ref{tab:coupled_qubit_params}. Those with varying detuning are listed in Table~\ref{tab:coupled_qubit_sweep}. We use $\phi_A=\phi_B$ for all simulations. \textbf{(a)} Gate schematic: microwave tones are applied simultaneously to the qubits off-resonantly via individual RF flux lines will induce a dynamical $ZZ$ coupling that entangles the qubits. \textbf{(b)} Dynamical $ZZ$ rate $\zeta_{ZZ}$ with varying drive frequency $\omega_d$ for drive amplitudes $\epsilon_A/2\pi=\epsilon_B/2\pi=10~\mathrm{MHz}$. \textbf{(c)} $ZZ$ rate for varying drive frequency $\omega_d$ and amplitudes $\epsilon_A = \epsilon_B$. \textbf{(d)} $ZZ$ rate with varying qubit-qubit detuning for microwave drives applied at $\omega_d/2\pi= \omega_B/2\pi+50~\mathrm{MHz}$. Three sets of amplitudes are chosen: \{[$\epsilon_A = 2\epsilon_B$,  $\epsilon_B/2\pi=50~\mathrm{MHz}$], [$\epsilon_A = 2\epsilon_B$,  $\epsilon_B/2\pi=100~\mathrm{MHz}$], and [$\epsilon_A = 4\epsilon_B$,  $\epsilon_B/2\pi=100~\mathrm{MHz}$]\}. \textbf{(e)} Gate error with varying cosine pulse length $\tau_g$, for microwave drives applied at $\omega_d = \omega_B+\Delta_B$, $\Delta_B/2\pi=\{-60, -30, 25, 50\}~\mathrm{MHz}$. \textbf{(f)} Gate error for varying qubit-qubit detuning and different gate time $\tau_g = \{100,200,300\} ~\mathrm{ns}$. We use flat-top cosine pulses with ramping time $\tau_\mathrm{ramp}=\tau_g/2$ for $\tau_g=100~\mathrm{ns}$, and $\tau_\mathrm{ramp}=\tau_g/3$ for the other gate times. The microwave drives are applied at $\omega_d/2\pi= \omega_B/2\pi+50~\mathrm{MHz}$.}
    \end{figure*}
    
    Another useful two-qubit gate is the CZ gate, which is equivalent to the CNOT gate up to single-qubit rotations. Several microwave-activated CZ gates in superconducting circuit architectures have been implemented, including those utilizing the higher levels in transmons \cite{chow2013microwave,krinner2020demonstration} or fluxonium \cite{ficheux2021fast,xiong2021arbitrary}, resonator-induced phase gate \cite{paik2016experimental}, parametric gate enabled by second order nonlinearity \cite{noguchi2020fast}, and recently based on differential AC-Stark shift of the computational states \cite{mitchell2021hardware,wei2021quantum}.
    
    Here, we explore the CZ gate scheme where the microwave pulses are applied off-resonantly to the low-frequency qubits, as depicted in Fig.~\ref{fig7}(a). We note that since the static $ZZ$ rate is negated via the multi-path coupling, the $ZZ$ interaction discussed in this section is purely dynamical. When the computational states are dressed by the microwave drives with respective amplitudes $\epsilon_{\{A,B\}}$ (which have cyclic frequency unit here), phases $\phi_{\{A,B\}}$, and at a frequency $\omega_d$ detuned from the qubit frequencies by $\delta_{\{A,B\}}=\omega_d - \omega_{\{A,B\}}$, a finite qubit-qubit exchange coupling $J_\mathrm{eff}$ turns on the dynamical $ZZ$ interaction \cite{mitchell2021hardware,wei2021quantum},
    \begin{equation}
    \label{eqn:dynamical_ZZ_analytical}
        \zeta_{ZZ}\approx 4\pi J_\mathrm{eff}\frac{\epsilon_A \epsilon_B}{\delta_A \delta_B} \cos(\phi_A - \phi_B),
    \end{equation}
    which can be used in combination with single-qubit gates $\hat{I}\hat{Z}$ and $\hat{Z}\hat{I}$ to implement a CZ gate following the relation
    \begin{equation}
    \label{eqn:CZfromZZ}
        \hat{U}_{CZ}=\exp\left[{\frac{i\pi}{4}(\hat{Z}\hat{I}+\hat{I}\hat{Z}-\hat{Z}\hat{Z}-\hat{I}\hat{I})}\right].
    \end{equation}
    
    We estimate the gate rate given by Eq.~(\ref{eqn:dynamical_ZZ_analytical}) and compare it with numerically obtained results for a pair of coupled fluxoniums with parameters listed in Table~\ref{tab:coupled_qubit_params}, together with coupling coefficients $J_L = 2~\mathrm{MHz}$ and $J_C = 11.5~\mathrm{MHz}$, corresponding to an effective spin-spin exchange interaction rate $J_\mathrm{eff} \approx 11~\mathrm{MHz}$. When applying two continuous microwave tones at a frequency $\omega_d$ with the same phases, $\phi_A = \phi_B$, and amplitudes $\epsilon_A=\epsilon_B$, we simulate the dynamical evolution of the system using the master equation, then compute the phase evolution as $\phi_{ZZ}(t) = \phi_{00}(t) + \phi_{11}(t) - \phi_{01}(t) - \phi_{10}(t)$. Fitting $\phi_{ZZ}(t)$ to a simple linear function of time yields the corresponding $ZZ$ rate.
    
    Figure~\ref{fig7}(b) shows excellent agreement between numerically and analytically obtained rates for relatively weak drive amplitudes $\epsilon_A/2\pi=\epsilon_B/2\pi=10~\mathrm{MHz}$. Notably, the analytical solution diverges when the perturbation condition $\epsilon_{A,B}\ll |\delta_{A,B}|$ is violated, specifically when the drive is close the qubits' frequencies. Meanwhile, the numerical result shows high, but finite, $ZZ$ rate in these regions.
    
    To further explore possible gate rates, we perform a two-dimensional sweep as displayed in Fig.~\ref{fig7}(c). An induced $ZZ$ rate of over $50$ MHz can be reached for $\epsilon_A/2\pi = \epsilon_B/2\pi \geq 100~\mathrm{MHz}$, and large drive-qubit detuning can be compensated by higher amplitudes to produce large $ZZ$ rate. Since we can pick the drive frequency independently from the qubit frequencies, this gate scheme is more versatile compared to the previously discussed CR gate.
    
    It is important to investigate gate performance for large qubit-qubit detunings upon scaling up. To this end, we fix qubit A's parameters and vary qubit B's inductive energy as shown in Table~\ref{tab:coupled_qubit_sweep}, apply the microwave drive at frequency $\omega_d/2\pi = \omega_B/2\pi + 50~\mathrm{MHz}$, then compute the dynamical $ZZ$ rate for varying qubit-qubit detuning $\Delta=\omega_B - \omega_A$. Interestingly, the dynamical $ZZ$ rate shown in Fig.~\ref{fig7}(d) is quite high at large detuning $\Delta/2\pi = 900~\mathrm{MHz}$, with $|\zeta_{ZZ}|/2\pi > 1~\mathrm {MHz}$ for $\epsilon_A/2\pi = 200~\mathrm{MHz}$, $\epsilon_B/2\pi = 100~\mathrm{MHz}$. This promises short gate time $\tau_g$ even for qubits far detuned from each other.
    
    We proceed to simulate the gate performance by calibrating the appropriate gate parameters and computing the fidelity following Eq.~(\ref{eqn:two_q_fidelity}). In the first step, we estimate the required amplitudes based on the conditional metric $R$ given by Eq.~(\ref{eqn:conditional}), with the qubits now initialized in the superposition state $|+\rangle_A|+\rangle_B$, where $|+\rangle_A=(|0\rangle_A+|1\rangle_A)/\sqrt{2}$. Since the drive frequency is also an independent parameter in this gate scheme, there are more variables to consider. Thus, in order to simplify the calibration, we first simply use a cosine pulse shape as defined by Eq.~(\ref{eqn:cosine_pulse}), sweep the gate time $\tau_g$, and vary the drive frequency $\omega_d$. We employ the Nelder-Mead optimization method with the drive amplitudes $\epsilon_{\{A,B\}}$ as free parameters to estimate the best gate fidelity.
    
    For a pair of coupled fluxoniums with parameters listed in Table.~\ref{tab:coupled_qubit_params} and coupling coefficients $\{J_L,J_C\}=\{2,11.5\}~\mathrm{MHz}$, the gate error $1-\mathcal{F}$ generally decreases with longer gate time $\tau_g$ (and correspondingly longer ramping time), as shown in Fig.~\ref{fig7}(e). Notably, larger detunings between the drive and the qubit generally correspond to lower gate error. From these results, we attribute the residual gate error to non-commuting single-qubit rotations such as $\hat{X}\hat{I}$ and $\hat{I}\hat{X}$ due to the strong drives. Interestingly, this implies that optimizing the drive frequency and the ramping time is important if high-fidelity operation is desired. For example, a 100-ns-long gate has a fidelity as low as $\sim 10^{-4}$ when the pulses are applied at a frequency $\Delta_B/2\pi = 50~\mathrm{MHz}$ above qubit B.
    
    Finally, we optimize the pulse parameters and evaluate the gate errors for varying qubit-qubit detuning $\Delta$ using pulses applied at a frequency 50 MHz above qubit B. For gate time $\tau_g=100~\mathrm{ns}$, we use the ramping time $\tau_\mathrm{ramp}=\tau_g/2=50~\mathrm{ns}$, and for $\tau_g=200,300~\mathrm{ns}$, we use $\tau_\mathrm{ramp}=\tau_g/3$. The results are shown in Fig.~\ref{fig7}(f), with the pulse amplitudes optimized using the Nelder-Mead method at each $\Delta$. For $\tau_g=100~\mathrm{ns}$, we observe an increase in gate error at higher $\Delta$, which can be explained by the lower gate rate, similar to the CR gate case. However, the gate error remains below $10^{-3}$ up to $\Delta/2\pi = 900~\mathrm{MHz}$ for 300-ns-long pulses.
    
    A notable source of gate error is finite leakage to higher states due to high-order multi-photon transitions, which manifests as the peaks in Fig.~\ref{fig7}(f) (see Appendix~\ref{appendix:leakage_error}). However, since this is typically caused by very large drive amplitudes, the gate error remains below $10^{-2}$. It can also be further suppressed by using pulses with lower amplitudes at the cost of longer gate times, as evidenced by the difference between the errors of 200-ns and 300-ns gates. Thus, we believe that this leakage does not degrade the gate performance significantly. In addition, choosing another drive frequency for the specific qubit parameters should improve the fidelity substantially.
    
    The small coherent gate errors suggest that the fidelity may be limited by decoherence processes. Since the quantum dynamics during the gate only involves the high-coherence computational states, coherence-limited error is expected to be small. As discussed in Appendix~\ref{appendix:decoherence_error}, the decoherence-induced error of a 300-ns two-qubit gate is below $10^{-3}$ for relaxation time $T_1>300~\mathrm{\mu s}$ and coherence time $T_2=2T_1$. For a large-scale fluxonium quantum processor, we expect the average two-qubit gate time to be $\sim 200$ ns, and average relaxation time to be $\sim 500~\mathrm{\mu s}$, resulting in a small average gate error of $1-\mathcal{F}=3\times 10^{-4}$.
    
    One potential hurdle for this type of off-resonant driven interaction is the degradation of coherence times as the drive amplitude increases \cite{wei2021quantum,mitchell2021hardware}, which can be attributed to either unstable electronics or possible interaction with two-level defects \cite{schneider2018local,carroll2021dynamics}. In the former case, since the qubits are controlled via direct pulses from high-precision AWGs, fluctuation due to IQ mixing would be reduced significantly. In the latter case, the spectral density of two-level defects in the proposed operating regime has not been rigorously studied, but is likely lower than at 5 GHz range \cite{pop2014coherent,nguyen2019high}. Future work on two-level system defects spectroscopy \cite{schneider2018local,carroll2021dynamics} in this frequency range will be needed to confirm the preliminary assessment.

\section{\label{sec:scaling}Frequency allocation}

A major challenge in engineering superconducting architectures, especially fixed-frequency platforms, is spectral crowding where fluctuations in qubit parameters lead to frequency collisions that degrade the performance of multi-qubit devices and subsequently nanofabrication yield \cite{brink2018device}. Our proposed architecture is constructed from fluxonium circuits biased at a fixed external flux to reach high coherence, so allocating the qubit frequencies in a quantum processor is of central importance in scaling up the platform. In this section, we discuss frequency fluctuation of the computational transition in fluxonium and experimentally feasible dispersion. Based on the gate results in previous sections, we impose a set of frequency constraints and simulate the yield of collision-free large-scale devices.

 \subsection{Frequency dispersion of computational states}
 
 \begin{figure}[t!]
    \includegraphics[width=0.4\textwidth]{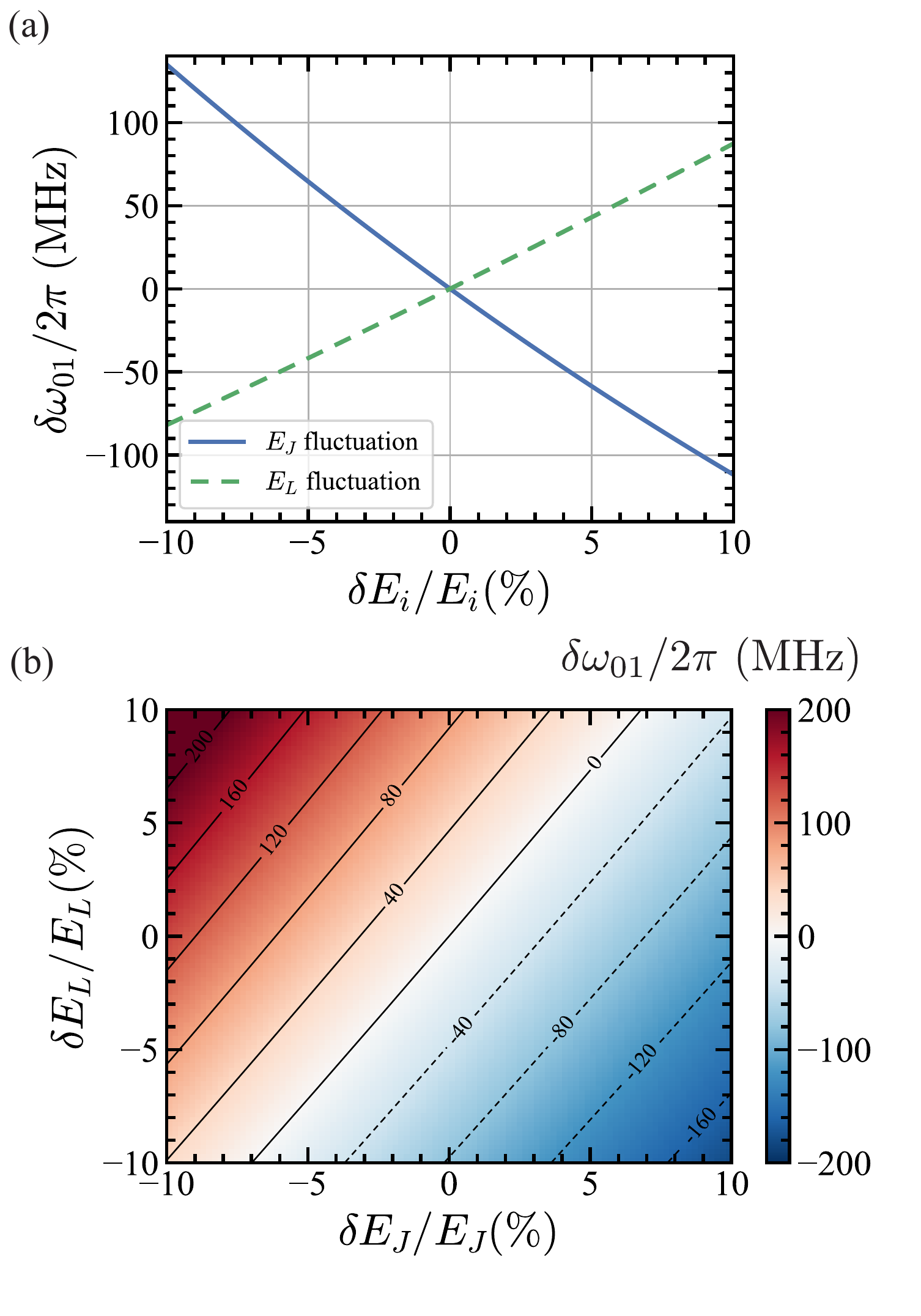}
    \caption{\label{fig9}Frequency dispersion for a fluxonium qubit with parameters $\{E_J,E_C,E_L\}=\{4,1,1\}~\mathrm{GHz}$. \textbf{(a)} $|0\rangle \rightarrow |1\rangle$ frequency variations resulted from changes in $E_L$ or $E_J$. \textbf{(b)} Frequency dispersion for both $E_L$ and $E_J$ fluctuations.}
\end{figure}
    
    Superconducting qubit frequencies depend on the tunneling barrier in Josephson junctions, thus are prone to fluctuation. The degree of randomness is typically characterized by the frequency dispersion $\sigma_f$. Although fluxonium circuit consists of more components compared to transmon, we argue that its frequency dispersion is quite small, so it should scale up more favorably. We treat $E_C$ as a fixed parameter since the capacitive shunt is highly reproducible with modern microwave design and engineering. This has been validated by recent results in transmon studies \cite{kreikebaum2020improving,hertzberg2021laser}. Since the qubit parameters result in low computational transition frequencies, fluctuations in $E_J$ and $E_L$ correspond to rather small differences in $|0\rangle\rightarrow |1\rangle$ frequencies, as shown in Fig.~\ref{fig9}. 
    
    Typical junction variations on a device correlate with the positions of the qubits, as the electron beam angle is not consistent across a big wafer \cite{kreikebaum2020improving}, or because the resist thickness is not uniform. In case the superinductor is made from a junction array, this results in $E_J$ and $E_L$ changing in the same way, compensating each other, such that a $2\%$ increase in both parameters corresponds to a frequency shift of only $\sim 10~\mathrm{MHz}$. In addition, junction aging also has the same effect on both $E_J$ and $E_L$. Hence, systematic frequency variation would be suppressed in fluxoniums.
    
    For random fluctuations of the oxide layer that cause unpredictable changes in $E_J$ and $E_L$, we consider the parameters separately. With $E_J$, we can employ a laser annealing technique that has been demonstrated to improve the nanofabrication precision of Josephson junctions to a dispersion of $0.3\%$ \cite{hertzberg2021laser,zhang2020high}. This corresponds to a frequency variation $\sigma_f\lesssim 10~\mathrm{MHz}$ when only $E_J$ changes. For $E_L$, if the superinductor is constructed from an array consisting of $N= 100$ junctions \cite{manucharyan2009fluxonium,masluk2012microwave,pop2014coherent}, independent fluctuations of individual junctions would reduce the fluctuation in $E_L$ by $\sqrt{N}$ times, or $\delta E_L /E_L \sim 0.2\%$ if each junction fluctuates by $2\%$. In addition, geometric superinductors have recently been demonstrated to have variation as low as $0.2\%$ as well \cite{peruzzo2021geometric}. This also corresponds to a frequency dispersion $\sigma_f\lesssim 10~\mathrm{MHz}$. Therefore, recent advances in Josephson junction fabrication techniques can be adapted to produce fluxonium devices with frequency dispersion in the range $\sigma_f\sim 10~\mathrm{MHz}$, assuming the worst case of random fluctuations in both $E_J$ and $E_L$.
    
\subsection{Frequency constraints \& fabrication yield}

Spectral crowding leads to frequency collisions of neighboring qubits where control of one qubit affects others. This leads to lower probability of successful fabrication of a good multi-qubit device. In the field of nanofabrication, this probability is defined as \textit{yield}. We analyze the yield of our proposed architecture with square lattice topology in Fig.~\ref{fig0}, which has a high number of qubit-qubit connections. Other topologies such as heavy square or heavy hexagon that are used to implement hybrid quantum error correction codes \cite{chamberland2020topological} would have fewer connections and subsequently higher yield \cite{hertzberg2021laser,morvan2021optimizing}.

To avoid collisions, we impose the following constraints on the qubit frequencies, as summarized in Table~\ref{table:freq_allo}. Due to the high anharmonicity of fluxonium in the proposed regime, we only consider the frequencies of computational states.

    \begin{table}[b!]
    \caption{Frequency collision constraints and definitions. For spectator CR, the drive frequency is on resonance with the target qubit, $\omega_d=\omega_j$.}
    \label{table:freq_allo}
        \begin{tabular}{ |c|c|c|  }
         \hline
         Constraint & Definition \\
         \hline
         \rule{0pt}{2ex} High coherence & $0.2~\mathrm{GHz} \leq \omega_i/2\pi \leq 1.2~\mathrm{GHz}$\\
         \hline 
         \rule{0pt}{2ex} Addressability &  $20~\mathrm{MHz} \leq |\omega_i-\omega_j|/2\pi$  \\
         \, & $10~\mathrm{MHz} \leq|2\omega_i-\omega_j|/2\pi$\\
         \hline
         \rule{0pt}{2ex} CR gate & $20~\mathrm{MHz} \leq |\omega_i-\omega_j|/2\pi\leq 1~\mathrm{GHz}$\\
         
         CZ gate & $20~\mathrm{MHz} \leq |\omega_i-\omega_j|/2\pi\leq 1~\mathrm{GHz}$\\
         \, & $10~\mathrm{MHz} \leq |\omega_d-(\omega_i+\omega_j)/2|/2\pi$\\
         \hline
         
         Spectator & $ 20~\mathrm{MHz}\leq|\omega_d-\omega_k|/2\pi$ \\ 
         \, & $10~\mathrm{MHz}\leq|2\omega_d-\omega_k|/2\pi$ \\
         \hline 
        \end{tabular}
    \end{table}

    -\emph{High coherence}: For the qubit to be in the high coherence regime, we target specifically the frequency range from $0.2$ to $1.2$ GHz. The high relaxation time $T_1$ is achieved here mainly due to the suppression of dielectric loss at low frequency \cite{nguyen2019high,zhang2020fast}.
    
    -\emph{Addressability}:  To avoid having microwave pulses applied on-resonant to qubit $i$ inducing unwanted rotation in a nearest neighbor qubit $j$ \cite{kelly2014optimal,sung2021realization}, we require their $|0\rangle\rightarrow|1\rangle$ transition frequencies to be separated by $|\omega_i - \omega_j|/2\pi \geq 20~\mathrm{MHz}$. In addition, we impose an additional constraint to avoid two-photon driving, $|2\omega_i - \omega_j|/2\pi \geq 10~\mathrm{MHz}$. This frequency separation is lower since multi-photon processes require higher drive amplitudes. We note that these values are based on previous transmon studies \cite{hertzberg2021laser}, and future experiment on planar fluxonium devices will be needed to pinpoint the lower bound of the constraint. 
    
    -\emph{High-fidelity entangling gates}: As discussed previously, gate time and fidelity of entangling operations depend on the detuning between the qubit frequencies. Here, we target two-qubit gate error to be lower than $10^{-2}$. To this end, we rely on the gate simulations in Sec.~\ref{sec:two_qubit_gates} and restrict the detuning between the participating qubits' frequencies $|\omega_i - \omega_j|/2\pi$ to be between [20 MHz - 1 GHz] for both gate schemes. In addition, to avoid inducing two-photon $\mathrm{SWAP}$-like gate \cite{nesterov2021proposal}, we exclude the frequency region $(\omega_i + \omega_j)/2$ from the possible region for driving frequency $\omega_d$. Future work on pulse optimization and advanced calibration of gate parameters may reduce the error further, relaxing these restrictions. In such a case, our proposed range can simply be extended to give higher yields.

    -\emph{Spectator error}: Beyond interaction with qubit $j$ which participates in an entangling operation, two-qubit gate pulses applied to qubit $i$ can also induce unwanted transitions in a connected qubit $k$ if the drive frequency $\omega_d$ is close to its transition frequency $\omega_k$. Since qubit $k$ does not affect the intended operation, this is known as \textit{spectator} error. To avoid collision with spectator qubits, we impose the additional constraints on the drive frequency such that it is sufficiently detuned from the spectating qubits' frequencies, $|\omega_d - \omega_k|/2\pi\geq 20~\mathrm{MHz}$ and, including two-photon driving, $|2\omega_d - \omega_k|/2\pi\geq 10~\mathrm{MHz}$. For the cross-resonance gate, the drive frequency is fixed at the target qubit frequency, $\omega_d = \omega_j$.
    
\begin{figure}
        \includegraphics[width=0.48\textwidth]{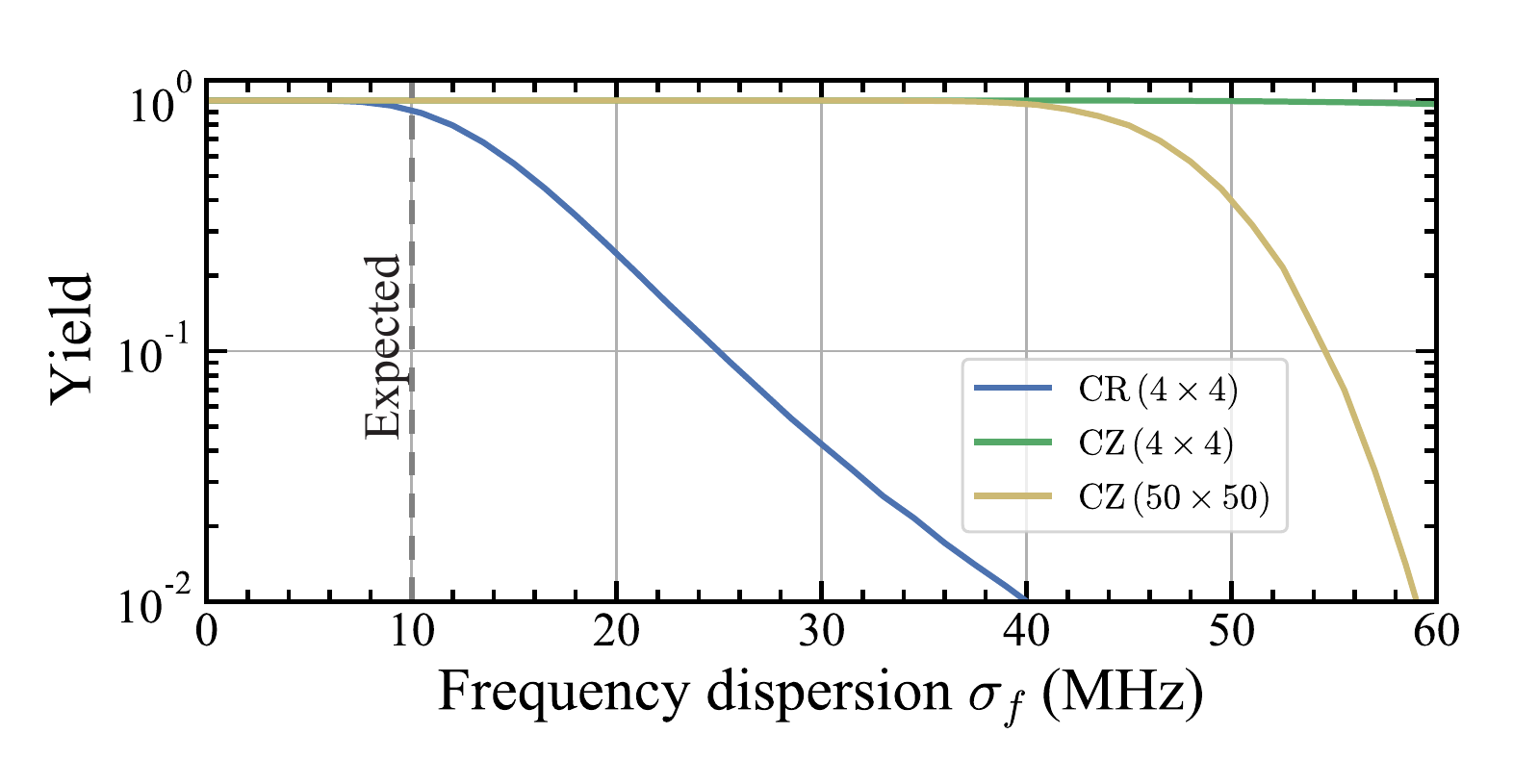}
        \caption{\label{fig8} Fabrication yield with varying qubit frequency dispersion $\sigma_f$ for square lattice topology based on cross-resonance (labeled CR) and differential AC-Stark (labeled CZ) two-qubit gates, with cells consisting of $4\times 4$ and $50\times 50$ qubits. As discussed in the main text, we expect a frequency dispersion of $\sigma_f\sim 10~\mathrm{MHz}$ if state-of-the-art nanofabrication techniques are employed to construct the processor.}
\end{figure}

The frequency allocation simulation is performed following Ref.~\cite{morvan2021optimizing}. We specify a directed graph based on the square lattice topology shown in Fig.~\ref{fig0}. Each qubit is defined as a node and each pair of qubits is connected via an edge. The orientation of the edge is important for the CR gate, but not for the AC-Stark CZ gate. Then, we find a set of frequencies on these nodes satisfying the conditions defined in Table~\ref{table:freq_allo}. Solution optimization is based on mixed-integer programming, and we use the Gurobi solver with Pyomo python package \cite{hart2011pyomo,hart2017pyomo}.

We show the yield results based on various qubit frequency dispersion $\sigma_f$ in Fig.~\ref{fig8}. For the square lattice topology, the yield based on AC-Stark effect is considerably higher than one based on CR effect since the former does not have a strict constraint on the drive frequency. Even for a large $50\times 50$ cell, the yield is close to unity for frequency dispersion $\sigma_f\leq 40~\mathrm{MHz}$. 

As the number of qubits in a quantum processor increases, individually optimizing all the frequencies on the chip is resource-heavy. Instead, an optimized unit cell is tiled to generate a larger lattice with periodic boundary condition. The corresponding yield estimation is given by
\begin{equation}
    y = y_\mathrm{cell}^{N_\mathrm{device}/N_\mathrm{cell}},
\end{equation}
where $y$ ($y_\mathrm{cell}$) is the yield of the large device (unit cell), and $N_\mathrm{device}$ ($N_\mathrm{cell}$) is the number of qubits in the device (unit cell). For example, based on the $50\times 50$ data, a device containing $10^4$ fluxonium qubits operated using AC-Stark CZ gate would have fabrication yield $y\approx 60\%$ if the average frequency dispersion is $\sigma_f=42~\mathrm{MHz}$, and $y\approx 100\%$ for $\sigma_f<30~\mathrm{MHz}$.  Thus, the combination of the flexibility of AC-Stark CZ gate and the high anharmonicity of fluxonium qubit promises a high-yield, scalable superconducting platform.



\section{\label{sec:qec}Toward Logical Qubits}
    \begin{figure}
    \centering
        \includegraphics[width=0.5\textwidth]{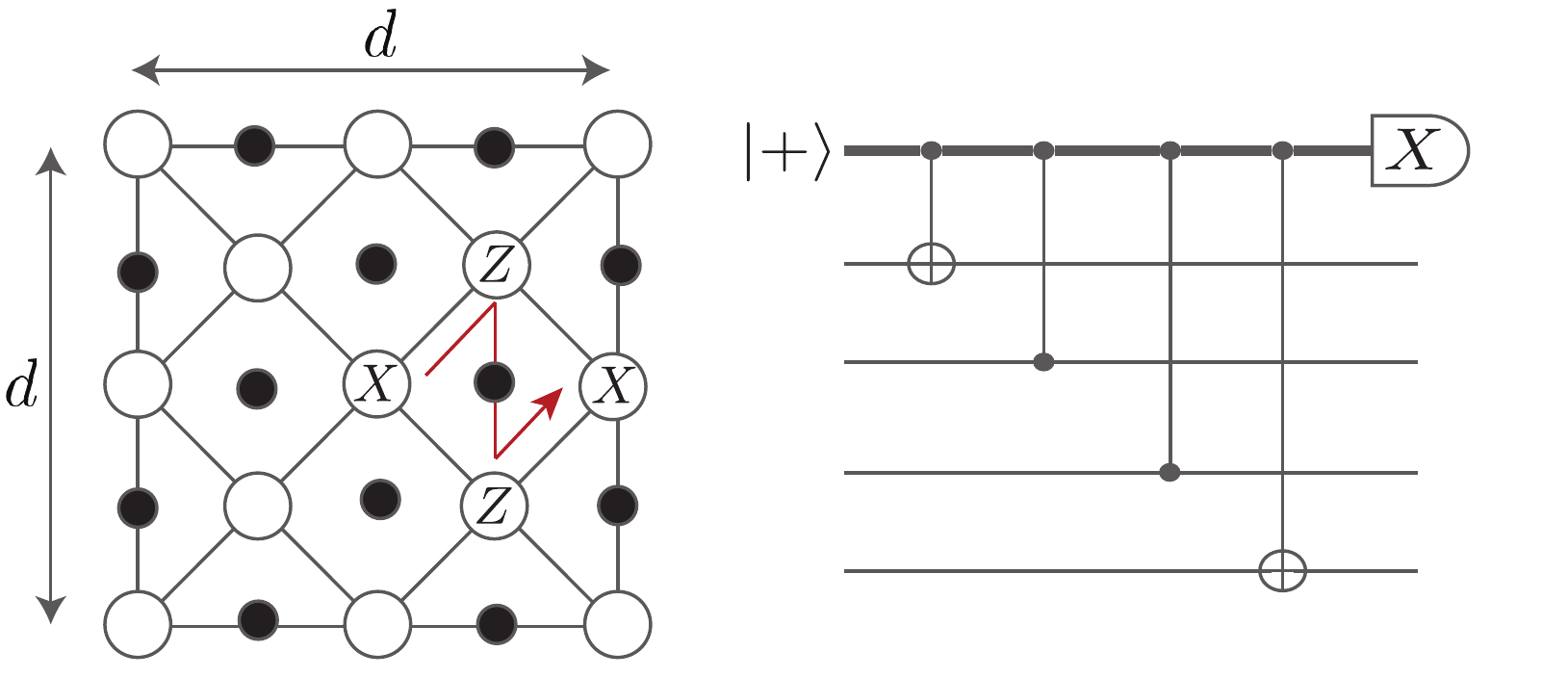}
    \caption{(Left) Illustration of a regular square XZZX code with distance $d$ where data (ancilla) qubits are shown as open (closed) circles. Each square stabilizer is the multi-qubit Pauli operator $X\otimes Z\otimes Z\otimes X$. The triangular stabilizers on the horizontal boundaries are $X\otimes Z\otimes X$, and the ones on left (right) vertical boundary are $Z\otimes Z\otimes X$ ($X\otimes Z\otimes Z$). The order in which the qubits are coupled to the ancilla at the center of each face is indicated by the red arrow. (Right) Stabilizer measurement circuit for the XZZX code. The ancilla is prepared in state $\ket{+}$, then coupled to data qubits with CNOT and CZ gates, and finally measured in the $X$ basis.}
    \label{fig:xzzx}
    \end{figure} 

Logical qubits encoded using a quantum error-correction (QEC) code can achieve arbitrarily low error rates as long as the physical error rate is below a certain threshold \cite{aharonov2008fault,knill1998resilient,preskill1998reliable}. The local connectivity and high thresholds make surface code \cite{bravyi1998quantum,fowler2009high,wang2011surface,fowler2012surface} a popular choice for realizing scalable quantum computation~\cite{kitaev2003fault,dennis2002topological}. In this section, we investigate the performance of the XZZX surface code~\cite{ataides2021xzzx} using fluxonium qubits, as illustrated in Fig.~\ref{fig:xzzx}. The figure shows the quantum circuit used to measure the stabilizers where an ancilla qubit, initialized in the superposition state $|+\rangle$, is placed at the center of each face. Next, a sequence of CNOT and CZ gates are applied, and finally the ancilla is measured in the X basis. For numerical simulations of the circuit shown in the figure, errors are applied before the gates, after preparation (or reset), before readout, and on idle qubits during these operations. 

\begingroup
\setlength{\tabcolsep}{10pt} 
\renewcommand{\arraystretch}{1.5}
\begin{table}[b]
    \caption{\label{tab:pauli_error}Expected average Pauli errors in the processor for different average relaxation times $T_1$.}
        \begin{tabular}{|l||c|c|c|} 
        \hline
          Opera-  & \multicolumn{3}{|c|}{Error $\varepsilon$}   \\
         [0.1ex] 
         \cline{2-4}
          tion & $300~\mathrm{\mu s}$ &$700~\mathrm{\mu s}$ &$1~\mathrm{m s}$ \\
         \hline\hline
         2Q CZ  & $5.3\times 10^{-4}$& $2.3\times 10^{-4}$& $1.6\times 10^{-4}$   \\ [0.1ex]
         \hline
         1Q H  & $1.1\times 10^{-5}$& $4.7\times 10^{-6}$& $3.3\times 10^{-6}$ \\[0.1ex]
         \hline
         Readout  & $10^{-2}$& $10^{-2}$& $10^{-2}$\\[0.1ex]
         \hline
         Reset & $10^{-2}$& $10^{-2}$& $10^{-2}$ \\ [0.1ex] 
         \hline
         Idle (2Q)  & $2.2\times10^{-4}$& $9.5\times10^{-5}$& $6.6\times10^{-5}$ \\[0.1ex]
         \hline 
         Idle (1Q)  & $1.1\times10^{-5}$& $4.7\times10^{-6}$& $3.3\times10^{-6}$ \\[0.1ex]
         \hline 
         Idle (R) & $2.2\times10^{-4}$& $9.5\times10^{-5}$& $6.6\times10^{-5}$ \\[0.1ex]
         \hline 
        \end{tabular}
    \end{table}
    
The error channel for each of these operations is parametrized using the corresponding physical error rate $\varepsilon$ listed in Table~\ref{tab:pauli_error}. We estimate the physical error budget as follows. The readout and reset error values are taken from state-of-the-art experimental implementation in Ref.~\cite{gusenkova2021quantum}. The measurement time of fluxonium using quantum-limited parametric amplifiers is 200 ns. Since the reset is performed using feedback, and in principle can be done by heralding \cite{johnson2012heralded}, we assume its fidelity is limited by the readout fidelity, and state initialization in the physical qubits can be done instantaneously. We expect the average single-qubit gate time on the chip to be $10$ ns, and two-qubit CZ gate time to be $200$ ns. From these assumptions, we can compute the decoherence-limited gate fidelity, given by Eq.~(\ref{eqn:decoherence_fidelity}) in Appendix~\ref{appendix:decoherence_error}. These are lower than the average coherent gate errors. We also include idle errors in non-parcicipating qubits due to decoherence during the operations of participating qubits. Non-Markovian noise sources such as leakage and crosstalk are not taken into account.

To model single-qubit noise, the errors are chosen uniformly at random from the set $\{X,Y,Z\}$, each with probability $\varepsilon/3$ such that the total error probability $\varepsilon$ corresponds to the values listed in Table~\ref{tab:pauli_error}. For the two qubit CZ gate, errors are randomly and uniformly chosen from the set $\{I,X,Y,Z\}^{\otimes 2}/(I,I)$ each with probability $\varepsilon/15$ so that $\varepsilon$ corresponds to the error probability for the CZ gate in Table~\ref{tab:pauli_error}. The CNOT gate is applied by sandwiching the CZ gate between two Hadamards, so the error channel for CNOT is composed of error channels of two Hadamards and one CZ. 

\begin{figure}[t]
        \includegraphics[width=0.47\textwidth]{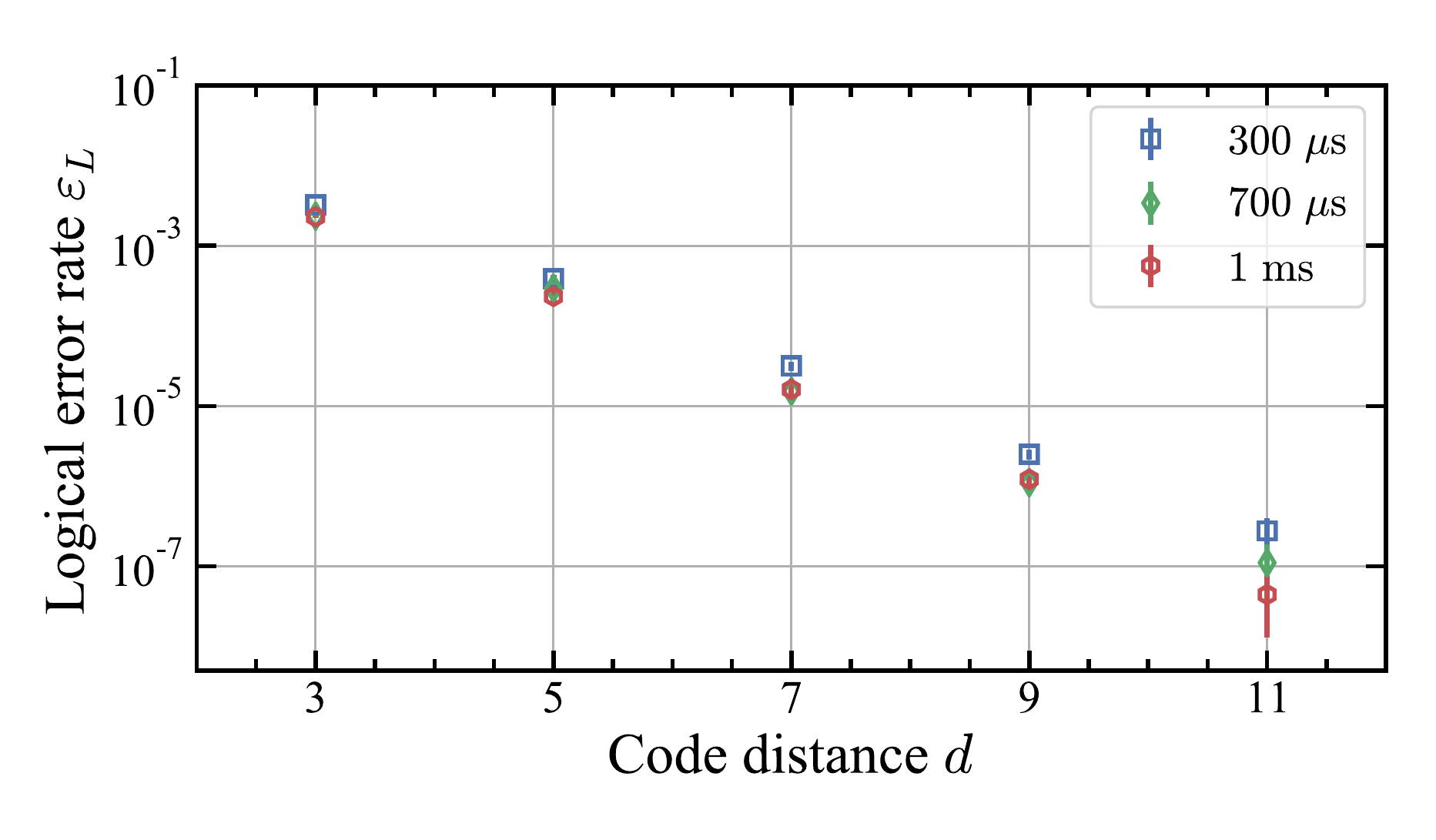}
        \caption{\label{fig11}Logical error rates for varying surface code distance $d$ with Pauli errors listed in Table~\ref{tab:pauli_error}. We perform Monte Carlo (MC) simulations involving $d$ rounds of faulty syndrome measurements using the circuit in Fig.~\ref{fig:xzzx}, on a $d\times d$ XZZX-code, also illustrated in the figure. In order to extract error rates in the range shown, $\{1.8\times 10^5, 1.8\times 10^5, 1.8\times 10^6, 1.8\times 10^7, 10^8\}$ MC simulations were performed for $d=\{3,5,7,9,11 \}$.}
\end{figure} 

We run Monte-Carlo (MC) simulations for a $d\times d$ XZZX-code for varying $d$, using the depolarizing noise described above for different qubit relaxation times $T_1$. In each MC simulation, all the stabilizers of the code are measured for $d$ rounds and error correction is performed using a minimum weight perfect matching decoder~\cite{dennis2002topological,edmonds1965paths,kolmogorov2009blossom,ataides2021xzzx}. Logical error rate is recorded as the ratio of number of MC simulations with a logical error and the total number of MC simulations. Fig.~\ref{fig11} shows the numerically obtained logical error rates $\varepsilon_L$ as a function of code distance $d$. We find that $\varepsilon_L$ decreases exponentially with $d$, which indicates that the physical error rates of these operations are below thresholds. In addition, the low logical error rate $\varepsilon_L\sim 10^{-7}$ at code distance $d=11$ promises reduced resource overhead upon scaling up.

Interestingly, Fig.~\ref{fig11} shows that the logical error rate $\varepsilon_L$ is largely independent of the relaxation time for small code distance $d$. This is because the dominant sources of noise in Table~\ref{tab:pauli_error} are ancilla readout and reset errors, $\varepsilon =1\%$, which do not depend on qubit relaxation time. This implies that readout error might be the limiting factor in future fluxonium quantum processors. Although numerically estimating the readout fidelity for fluxonium qubits is outside the scope of this work, we hope that our result will stimulate further research in this direction.

\section{\label{sec:summary}Summary}

In conclusion, we propose a novel architecture based on fluxonium qubits with excellent scaling potential. We show from first principles that the qubits, biased at half-integer flux quantum, would operate as small-footprint fixed-frequency quasi-two-level systems. They are measured dispersively using individual resonators coupled to a common bus, controlled in diplexing fashion, with single-qubit gates in the range of $10$ ns having errors below $10^{-6}$ across the computational frequency bandwidth of nearly 1 GHz. Both readout and control crosstalk are expected to be small. The reduced design complexity of these components promises higher resource efficiency in large-scale devices. The multi-path coupling approach negates the static $ZZ$ rate completely, while allowing a sufficient exchange coupling rate between the computational states. 

We numerically demonstrate fast, high-fidelity two-qubit entangling gates based on the cross-resonance and differential AC-Stark effects, with gate errors consistently below $10^{-2}$ for qubit-qubit detuning of up to 1 GHz, significantly relaxing the frequency allocation constraints in multi-qubit devices with high connectivity. Since the gate schemes are based on exchange interaction between high-coherence computational states, errors due to decoherence are expected to be low. Notably, our gate simulation is based on a nominal qubit-qubit exchange interaction rate and simple pulses. Future work to optimize the performance of these gates should further improve the fidelity under more stringent conditions, such as large qubit-qubit detuning.


After discussing possible routes to construct each element of the architecture, we explore its scalability by simulating the fabrication yield, relying on the range of qubit-qubit detuning that allows high-fidelity gates. Specifically, for a large-scale device consisting of ten thousands qubits arranged into a square lattice as shown in Fig.~\ref{fig0} and operated using the AC-Stark CZ gates, we estimate a yield close to unity for frequency dispersion expected from state-of-the-art nanofabrication technology. This allows scaling of the platform to a high number of physical qubits without sacrificing their performance, streamlining the implementation of practical quantum algorithms and quantum error correction codes. Assuming negligible errors due to crosstalk and leakage, we show the exponential suppression of logical error rate using the XZZX surface code. The result also reveals the importance of optimizing readout and initialization fidelity in fluxonium.

Upgrading to large-scale devices will likely involve innovative integration of the proposed components into a cross-pollination between fluxonium and flip-chip technology \cite{vahidpour2017superconducting,rosenberg20173d,kosen2021building,conner2021superconducting}. However, we note that near-term small-scale experiments utilizing charge driving or capacitive coupling in planar chips will still perform well at the expense of higher design and operational complexity. We believe that our results will on one hand stimulate further research and development efforts on fluxonium-based quantum architectures, and on the other hand motivate similar scalability studies of novel superconducting platforms such as the $\cos(2\hat{\varphi})$ \cite{ioffe2002possible,douccot2002pairing,douccot2003topological, gladchenko2009superconducting,bell2014protected,smith2020superconducting,smith2020magnifying}, the bi-fluxon \cite{kalashnikov2020bifluxon}, and the $0-\pi$ \cite{brooks2013protected,dempster2014understanding,groszkowski2018coherence,di2019control,gyenis2021experimental} qubits.

\section*{Acknowledgments}
We thank Fnu Setiawan, William P. Livingston, Ravi K. Naik, and Joachim Cohen for helpful discussions. L.B.N. is grateful for enlightening conversations with Vladimir Manucharyan. This work was supported by the Office of Advanced Scientific Computing Research, Testbeds for Science program, Office of Science of the U.S. Department of Energy under Contract No. DE-AC02-05CH11231. The numerical simulation was done using the \textit{Quantum Toolbox in Python} (QuTiP) software package \cite{johansson2012qutip,johansson2013qutip,li2021pulse}.

\section*{Appendices}
\appendix
\section{\label{appendix:fluxonium}Decoherence analysis}
\emph{Dielectric loss}: The energy relaxation rate for dielectric loss follows the Fermi golden rule as \cite{pop2014coherent,nguyen2019high,zhang2020fast}

    \begin{equation*}
            \Gamma_{01}^\mathrm{diel}(\omega_{01}) = \frac{\hbar \omega_{01}^2}{4E_CQ_\mathrm{diel}}|\langle 0|\hat{\varphi}|1 \rangle|^2 \left[\mathrm{coth}\left( \frac{\hbar\omega_{01}}{2k_BT} \right) +1\right].
    \end{equation*}
We extracted the effective dielectric loss quality factor from transmon and resonator experiments, $Q_\mathrm{diel}=5\times 10^{6}$, corresponding to a loss tangent of $\mathrm{tan}\delta_\mathrm{diel} = 2\times 10^{-7}$. For a transmon, this would translate to an energy relaxation time $T_1= Q/\omega_{01}\approx 160~\mathrm{\mu s}$ for $\omega_{01}/2\pi = 5~\mathrm{GHz}$. The effective temperature of the qubit is assumed to be $T=20~\mathrm{mK}$.

\emph{Quasiparticle Tunneling}: The energy relaxation rate for quasiparticle tunneling across a Josephson junction follows \cite{catelani2011quasiparticle,catelani2011relaxation}

\begin{equation}
        \Gamma_{01}^{\mathrm{qp}}(\omega_{01}) = \left|\langle 0 |\sin \frac{\hat{\varphi}}{2}  |1 \rangle \right| ^2 \times \frac{8E_J}{\pi\hbar} x_\mathrm{qp} \sqrt{\frac{2\Delta_\mathrm{Al}}{\hbar\omega_{01}}},
        \end{equation}
where $\Delta_\mathrm{Al}$ is the superconducting gap for aluminum, and $x_\mathrm{qp}$ is the quasiparticle density normalized by the density of Cooper pairs. 

Since the matrix element is zero for tunneling across the small junction at the symmetric flux bias \cite{pop2014coherent}, only the quasiparticles in the inductor affect energy relaxation. For an array of junctions, the rate can be summed up from individual tunneling events across each junction with index $\beta$ as \cite{vool2014non,nguyen2019high},

    \begin{equation}
    \label{eqn:relaxation_qp_multi}
    \begin{split}
    \Gamma_{01}^{\mathrm{qp}}(\omega_{01}) &= \sum_{\beta=0}^{M+1} \left| \langle 0| \sin \frac{\hat{\varphi}_\beta}{2} |1 \rangle \right|^2 \frac{8E_{J,\beta}}{\pi\hbar} x_\mathrm{qp} \sqrt{\frac{2\Delta}{\hbar\omega_{01}}}\\
    &\approx \left| \langle 0| \frac{\hat{\varphi}}{2} |1 \rangle \right|^2 \frac{8E_L}{\pi\hbar}x_\mathrm{qp}\sqrt{\frac{2\Delta}{\hbar\omega_{01}}}.
     \end{split}
    \end{equation}
    
    We use $x_\mathrm{qp}=5\times 10^{-9}$, which corresponds to $T_1=1~\mathrm{ms}$ at absolute temperature for a fluxonium qubit as reported in Ref.~\cite{somoroff2021millisecond}. Effect of temperature following detailed balance is included Ref.~\cite{glazman2021bogoliubov}, assuming qubit temperature $T= 20~\mathrm{mK}$.
    
    The energy relaxation rates are computed for different qubit parameters and added up to give the result shown in Fig.~\ref{fig1}(d).
    
    \emph{Thermal photon dephasing}: We follow the thermal photon dephasing rate given as \cite{rigetti2012superconducting}
    \begin{equation}
        \label{eqn:thermal_dephasing_exact}
        \Gamma^{\mathrm{th}}_\phi = \frac{\kappa_{\mathrm{tot}}}{2}\mathrm{Re} \left[ \sqrt{\left(1+\frac{i\chi_{01}}{\kappa_{\mathrm{tot}}} \right)^2 + \frac{4i\chi_{01} n_{\mathrm{th}}}{\kappa_{\mathrm{tot}}}} -1\right].
    \end{equation}
    In the simulation, we use resonator linewidth $\kappa_\mathrm{tot}/2\pi = 2~\mathrm{MHz}$ and resonator temperature $T=50~\mathrm{mK}$, which determines the average thermal photon number as $n_\mathrm{th}=[\exp(\hbar\omega_{R}/k_BT)-1]^{-1}$.
\section{\label{appendix:sgate}Single-qubit gate simulation}

\begin{figure}
        \includegraphics[width=0.45\textwidth]{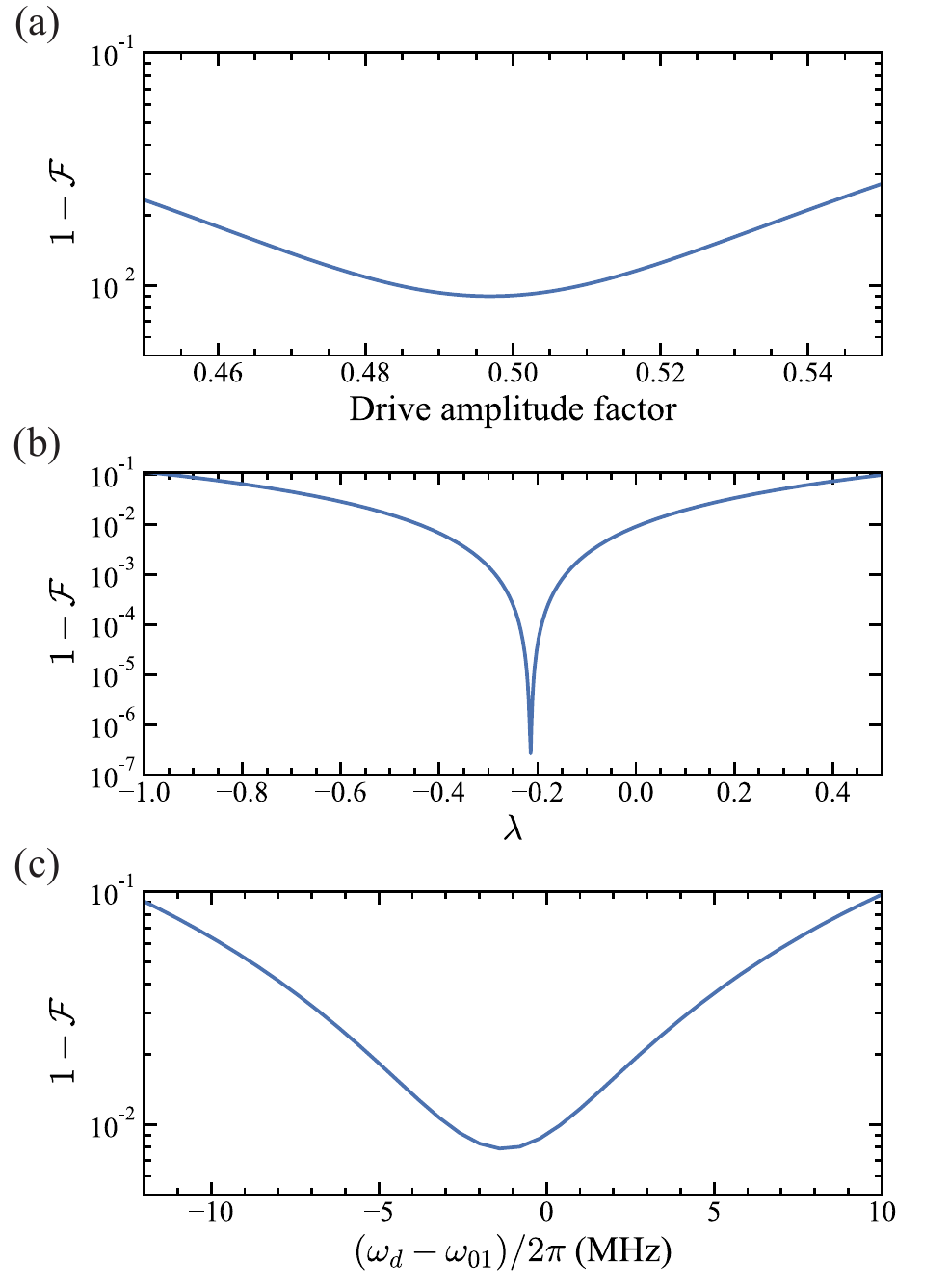}
        \caption{\label{figA_1Q_cal} Single-qubit gate fidelity simulation results for charge driving with $\tau_g=10~\mathrm{ns}$ and varying \textbf{(a)} relative drive amplitude factor, \textbf{(b)} DRAG coefficient $\lambda$, and \textbf{(c)} drive-qubit detuning. Each sweep is executed with other parameters set to default values: amplitude factor = 0.5,  $\lambda=0$, $\omega_d-\omega_{01}=0$.}
\end{figure}
We can estimate the required pulse amplitude to rotate the qubit by a certain angle as follows. When subjected to an on-resonant in-phase radiation tone with amplitude $\mathcal{E}_I(t)$, a two-level system with dipole moment $\eta$ undergoes a Rabi oscillation with frequency equal to $\Omega_r=\eta \mathcal{E}_I(t)$, assuming rotating wave approximation in the interaction frame. Thus, the qubit vector is rotated by an angle equal to $2\pi\int_{0}^{\tau_g}\eta \mathcal{E}_I(t)dt$ for gate time $\tau_g$. For a square pulse with constant amplitude $\mathcal{E}_I(t)=\epsilon_d$, a $2\pi$ rotation is realized when $\epsilon_d=(\eta\tau_g)^{-1}$, which we use to normalize the drive amplitude for other angles as well. In this work, we focus on $\pi$-rotations, corresponding to an amplitude factor of 0.5. For the cosine pulse defined in Eq.~\ref{eqn:cosine_pulse}, the time integration reads $\int_0^{\tau_g}\mathcal{E}(t)dt = \epsilon_d\tau_g/2$, so the amplitude condition for $2\pi$ rotation is $\epsilon_d=(\eta\tau_g/2)^{-1}$. From this relation, we note that the drive amplitude $\epsilon_d$ has unit of frequency.

To simulate a $\pi$ pulse, the single-qubit gate pulse is first set to these default values: amplitude factor = 0.5, DRAG coefficient $\lambda=0$, and detuning $\omega_d-\omega_{01}=0$. Ideally, this corresponds to a $\pi$ rotation that flips the qubit around $X$ or $Y$ axis, depending on the coupling degree of freedom (cf. Eq.~\ref{eqn:operators}). In practice, this leads to rotation error for short gate times as shown in Fig.~\ref{fig2}. 

To correct the effect from fast rotating terms, we attempt varying the relative amplitude of the pulse, then the DRAG coefficient $\lambda$, and finally the drive-qubit detuning. While keeping the other pulse parameters fixed, we compute the respective gate errors in each sweep. As shown in Fig.~\ref{figA_1Q_cal}, turning on a small negative DRAG coefficient suppresses most of the error, while the other approaches only suppress it within one order of magnitude. This validates the intuition that adding DRAG quadrature component would reshape the spectral profile of the pulse, improving the precision of the operation. To optimize the gate, we use all three parameters as free variables and minimize the error using Nelder-Mead method.

\section{\label{appendix:coupling}Coupled systems mapping}

\begin{figure}
        \includegraphics[width=0.4\textwidth]{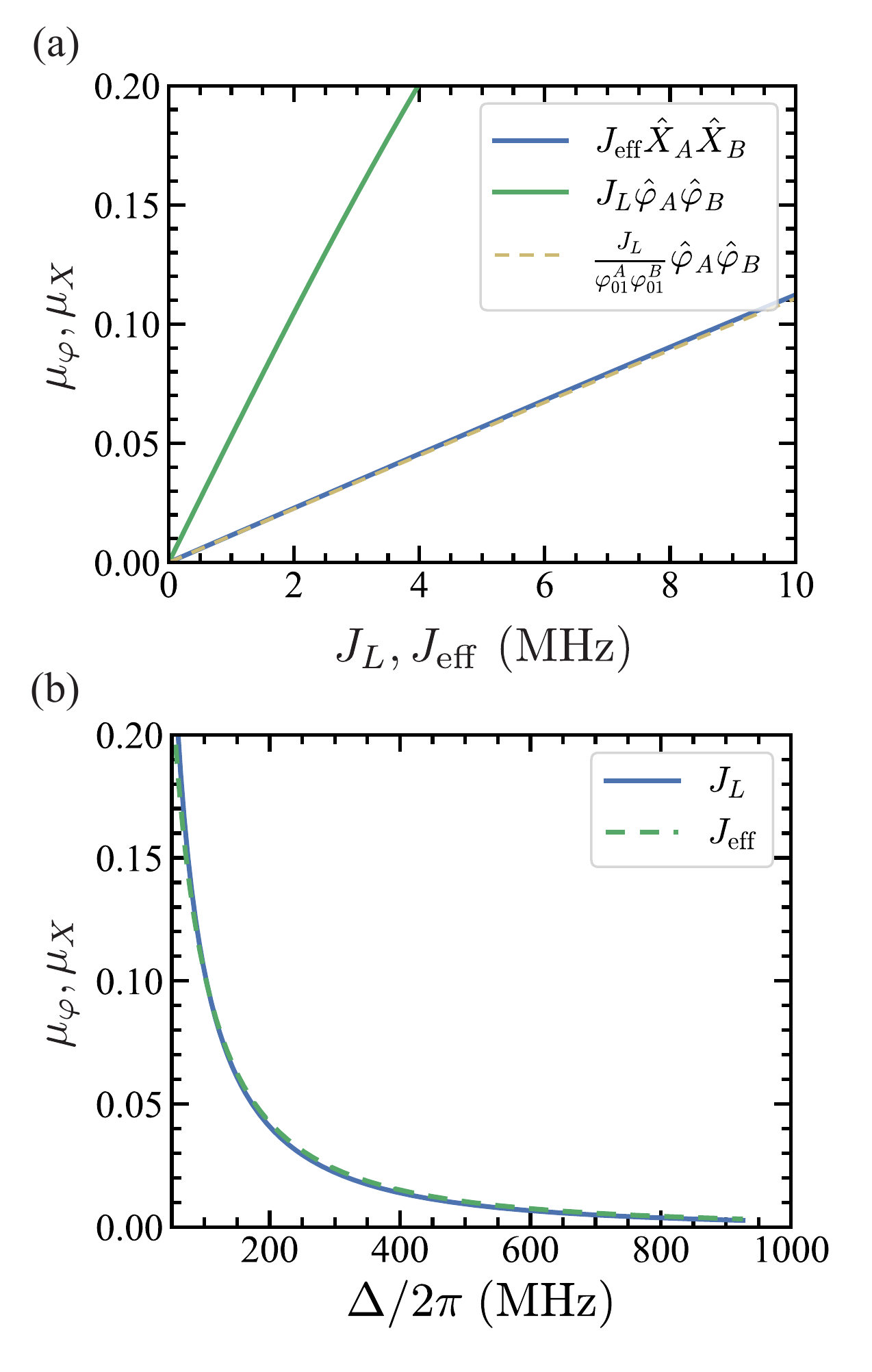}
        \caption{\label{figA_Jeff}Mapping between full model and simple model. \textbf{(a)} Level of mixing specified by $\mu_X$, $\mu_\varphi$ with respect to coupling coefficients $J_L$ and $J_\mathrm{eff}$. The pair of fluxoniums have a small capacitive coupling to cancel the static $ZZ$ rate. \textbf{(b)} Level of mixing specified by $\mu$ for varying qubit-qubit detuning as specified in Table~\ref{tab:coupled_qubit_sweep}. A small capacitive coupling $J_C$ is added at each detuning to cancel static $ZZ$.}
\end{figure}

We describe the mapping between an effective model of two coupled spins and a realistic system of two inductively coupled fluxonium qubits as follows. We sweep the coupling constants $J_\mathrm{eff}$ and $J_L$ in both models up to $10~\mathrm{MHz}$, and then compute the respective normalized cross matrix elements $\mu_X$, $\mu_\varphi$, where
\begin{equation*}
    \mu_X = \frac{\langle 00|\hat{X}_A\otimes \hat{I}_B|01 \rangle}{\langle 00|\hat{X}_A\otimes \hat{I}_B|10 \rangle},
\end{equation*}
\begin{equation*}
\mu_\varphi =  \frac{\langle 00|\hat{\varphi}_A\otimes \hat{I}_B|01 \rangle}{\langle 00|\hat{\varphi}_A\otimes \hat{I}_B|10 \rangle}.
\end{equation*}

Since $|\varphi_{01}|\approx 2.5$ for the proposed qubit parameters, the coupling term $\hat{H}_\mathrm{coupl}/h = -J_L\hat{\varphi}_A \hat{\varphi}_B$ in multi-fluxonium system produces more mixing for the same coupling coefficient compared to the coupled-spin model described by the Hamiltonian in Eq.~\ref{eqn:coupled_spins} with $\hat{H}_\mathrm{coupl}/h = J_\mathrm{eff}\hat{X}_A\hat{X}_B$, as shown in Fig.~\ref{figA_Jeff}(a). To match the level of mixing amplitude, we lower the coupling $J_L$ by a factor equivalent to $\varphi_{01}^A \varphi_{01}^B$, which produces a close match.

In other words, for linking the mixing level of a specific spin-spin coupling constant $J_\mathrm{eff}$, the required inductive coupling constant is $J_L\sim J_\mathrm{eff}/\varphi_{01}^A\varphi_{01}^B$. For example, to reach $J_\mathrm{eff}=10~\mathrm{MHz}$ in the coupled spin model, we need an inductive coupling $J_L \approx 2~\mathrm{MHz}$ in the coupled fluxonium system.

To explore the consistency of this mapping, we compute   $\mu_\varphi$ with $J_L \approx 2~\mathrm{MHz}$ and $\mu_X$ with $J_\mathrm{eff}=J_L \varphi_{01}^A\varphi_{01}^B$ for varying qubit parameters as listed in Table~\ref{tab:coupled_qubit_sweep}. To tune fluxonium B's frequency, $E_L$ is changed from 0.55 to 1.6 GHz while other parameters are fixed,
which corresponds to detuning $\Delta/2\pi$ from 27 to 926 MHz. The results in Fig.~\ref{figA_Jeff}(b) validate our mapping approach between the complex multi-level coupled fluxoniums system and the far simpler coupled spins model.

\section{\label{appendix:extended_error}Extended error budget}
\subsection{\label{appendix:decoherence_error}Decoherence errors}
With single- and two-qubit unitary error as low as $10^{-6}$, we turn to estimate the gate fidelity limited by decoherence. We assume that the qubit decays at rate $\Gamma_1$ and dephases at rate $\Gamma_\phi$, where these rates are related to the relaxation time $T_1$ and decoherence time $T_2$ as $\Gamma_1 = (T_1)^{-1}$, $\Gamma_\phi=(T_2)^{-1}-(2T_1)^{-1}$. A Pauli transfer matrix (PTM) of a single-qubit decoherence channel for duration $\tau$ is given as \cite{dawkins2020combining}
\begin{equation}
    \mathcal{E}(\tau) = 
    \begin{pmatrix}
    1 & 0 & 0 & 0 \\
    0 & e^{-(\Gamma_1/2+\Gamma_\phi)\tau} & 0 & 0 \\
    0 & 0 & e^{-(\Gamma_1/2+\Gamma_\phi)\tau} & 0 \\
    0 & 0 & 0 & e^{-\Gamma_1\tau} 
    \end{pmatrix}.
\end{equation}

\begin{figure}
    \includegraphics[width=0.47\textwidth]{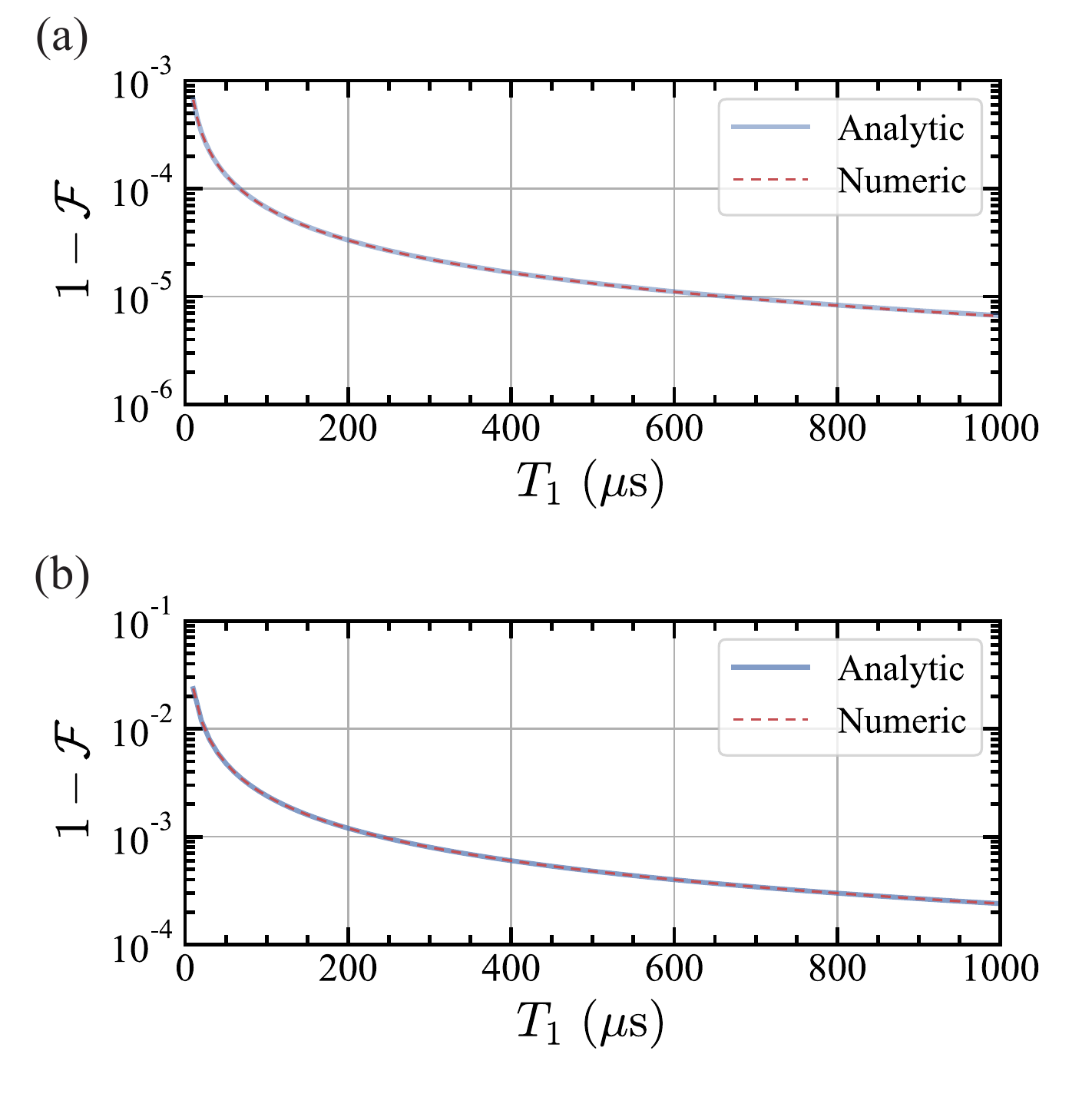}
    \caption{\label{fig_decohereceError}\textbf{(a)} Single-qubit and \textbf{(b)} two-qubit gate errors due to finite relaxation time $T_1$. The single-qubit pulse is 20-ns long, and the two-qubit pulses used to implement a CZ gate are 300-ns long.}
\end{figure}
    
In the absence of non-Markovian errors such as leakage and crosstalks, the PTM of a Pauli-twirled N-qubit channel is simply given by the tensor product $\mathcal{E}^{\otimes N}$. The process fidelity limited by decoherence can thus be written as
\begin{equation}\label{eqn:decoherence_fidelity}
\begin{split}
    \mathcal{F}_p &= \frac{1}{2^N}\mathrm{Tr}[\mathcal{E}^{\otimes N}] \\
    & = \frac{1}{2^N} \prod_{i=1}^N\left(1+e^{\Gamma_1^{(i)} \tau} + 2e^{-(\Gamma_1^{(i)}/2+\Gamma_2^{(i)}) \tau} \right),
\end{split}
\end{equation}
where $\Gamma_1^{(i)}$ and $\Gamma_\phi^{(i)}$ denote the energy relaxation and pure dephasing rates of qubit $i$, respectively.

To validate Eq.~\ref{eqn:decoherence_fidelity}, we simulate the process fidelity using the Lindblad master equation,
\begin{equation}
    \frac{d\hat{\rho}}{dt} = -i[\hat{H}_\mathrm{sys},\hat{\rho}] + \sum_\alpha \left(\hat{L}_\alpha\hat{\rho}\hat{L}_\alpha^\dagger -\frac{1}{2}[\hat{L}_\alpha^\dagger\hat{L}_\alpha, \hat{\rho}]\right),
\end{equation}
where $\hat{\rho}$ is the density matrix of the initial state, $\hat{H}_\mathrm{sys}$ is the Hamiltonian describing the evolution of the system without decoherence, and $\hat{L}_1 = \sqrt{\Gamma_1}|0\rangle\langle 1|$, $\hat{L}_\phi = \sqrt{\Gamma_\phi/2}(|0\rangle\langle 0| - |1\rangle\langle 1|)$ describes the relaxation and pure dephasing processes, respectively. Since only the computational subspace is involved in the proposed gate schemes in this work, we only take into account the decoherence of $|0\rangle$ and $|1\rangle$ states. To construct the process matrix for single- and two-qubit gates, we prepare $2^{2N}$ initial states, evolve them under $\hat{H}_\mathrm{sys}$, together with $\hat{L}_1 = \sqrt{\Gamma_1}|0\rangle\langle 1|$. We include only energy relaxation process since it is not clear what currently accounts for pure dephasing time in high-coherence fluxonium, and $T_2$ is primarily limited by $T_1$ \cite{nguyen2019high,zhang2020fast,somoroff2021millisecond}.

The dynamical simulation then gives us $2^{2N}$ output density matrices, which are then analyzed and, together with the input states, converted to a PTM $\mathcal{R}$ using maximum likelihood estimation method \cite{chow2012universal}. The process fidelity is defined as $\mathcal{F}_p = \mathrm{Tr}(\mathcal{R}_\mathrm{ideal}^\dagger \mathcal{R})/(2^{2N})$, from which we can compute the gate fidelity, $\mathcal{F}=(2^N\times \mathcal{F}_p+1)/(2^N+1)$. Figure~\ref{fig_decohereceError} shows the results from both analytical calculation and numerical simulation for a 20-ns-long single-qubit gate and a 300-ns-long CZ two-qubit gate. The perfect match between the analytical and numerical curves convinces us that Eq.~\ref{eqn:decoherence_fidelity} provides an excellent estimation of decoherence-limited fidelity. 

\subsection{\label{appendix:leakage_error}CZ gate leakage errors}
    
    \begin{figure}[t!]
            \includegraphics[width=0.45\textwidth]{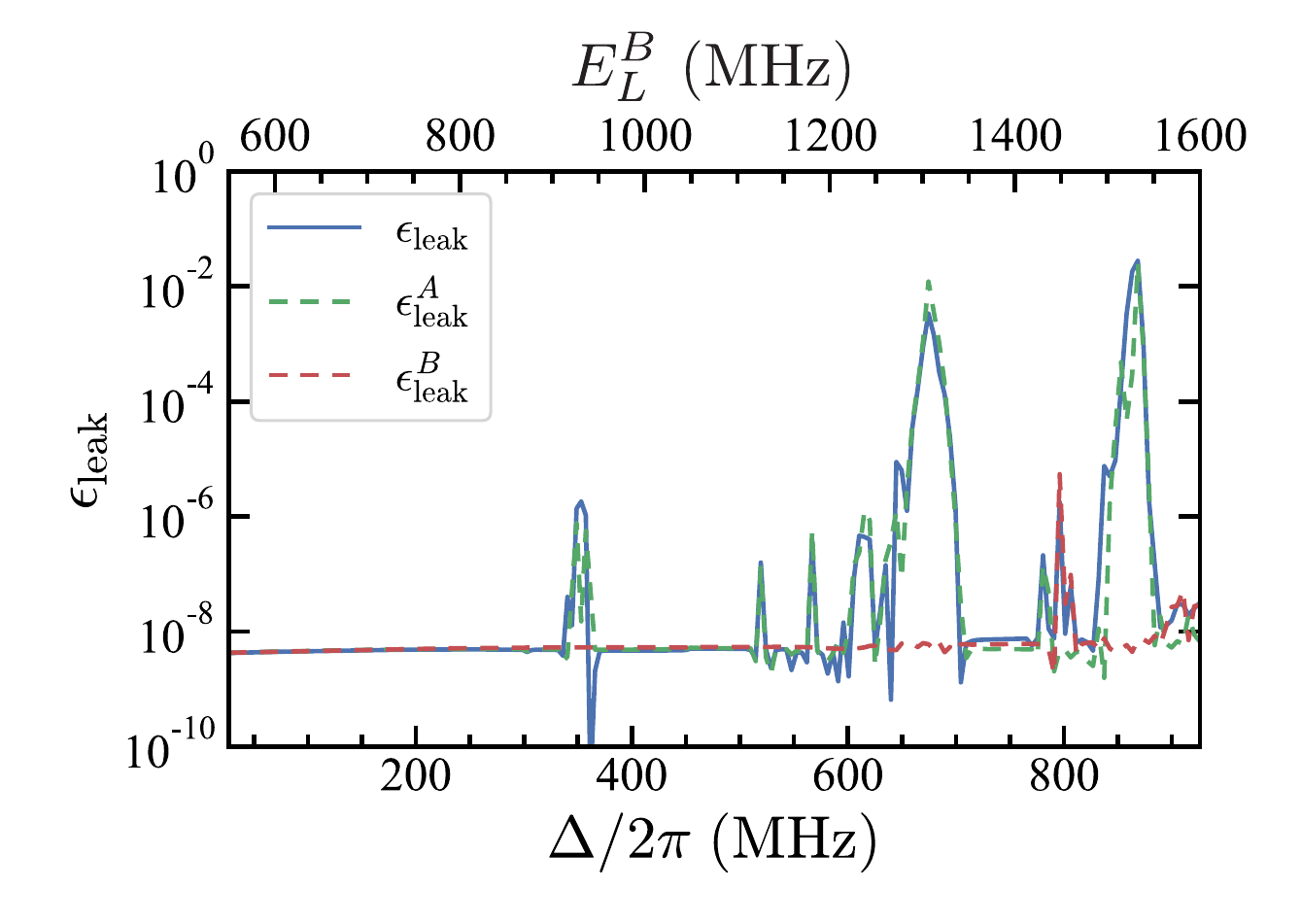}
            \caption{\label{fig_leakage} Leakage to higher states when microwave drives are used to implement a 200-ns CZ gate. The corresponding gate fidelity is plotted in Fig.~\ref{fig7}(f).}
    \end{figure}
    
As discussed in Sec.~\ref{sec:two_qubit_gates}, we observe spikes in the CZ gate errors at certain detuning $\Delta$, corresponding to specific drive frequency $\omega_d$. We attribute this to leakage outside of the computational subspace via high-order processes. To confirm this, we numerically compute the leakage defined as $1-(P_{|00\rangle}+P_{|01\rangle}+P_{|10\rangle}+P_{|11\rangle})$ using the optimal drive amplitudes given by the Nelder-Mead optimization, as discussed in Sec.~\ref{sec:two_qubit_gates}. 
    
In addition, we simulate the dynamics of each qubit when driven separately. Intuitively, since the drive frequency is far detuned from qubit A, the required drive amplitude for qubit A is much higher, $\epsilon_A\gg \epsilon_B$. Therefore, qubit A is more likely to have leakage. Indeed, this is confirmed by the results shown in Fig.~\ref{fig_leakage}. 

The most direct mitigation technique is to find a better drive frequency, which would help (i) reduce the required drive amplitude on qubit A to induce the necessary $ZZ$ rate, and (ii) avoid the bad frequency region. Nevertheless, the gate fidelity is still high when this high-order leakage is present, with $\epsilon_\mathrm{leak}\approx 10^{-2}$ in the worst case. Another approach is to use a longer pulse, at the cost of having error due to decoherence. Since the coherence time of the qubit can be in the range of millisecond, a good compromise can be made between the two sources of errors to have good gate fidelity following this approach.

\bibliography{apssamp}

\end{document}